\documentclass[nofootinbib,eprint,aps,twocolumn,british,  superscriptaddress]{revtex4-1}
\usepackage[LGR,T1]{fontenc}
\usepackage[latin9]{inputenc}
\usepackage{color}
\usepackage{amsmath}
\usepackage{amssymb}
\usepackage{graphicx}
\usepackage{wasysym}
\usepackage[caption=false]{subfig}

\makeatletter

\newcommand{\q}[1]{``#1''}

\ProvideTextCommand{\~}{LGR}[1]{\char126#1}

\providecommand{\tabularnewline}{\\}

\@ifundefined{showcaptionsetup}{}{ \PassOptionsToPackage{caption=false}{subfig}}
\usepackage[caption=false]{subfig}
\usepackage{subfig}
\makeatother

\usepackage{babel}

\begin{document}

\title{Efficient nonparametric $\boldsymbol{n}$-body force fields from machine learning}

\author{Aldo Glielmo}
\email{aldo.glielmo@kcl.ac.uk}
\affiliation{Department of Physics, King's College London, Strand, London WC2R 2LS, United Kingdom}
\author{Claudio Zeni} 
\email{claudio.zeni@kcl.ac.uk}
\affiliation{Department of Physics, King's College London, Strand, London WC2R 2LS, United Kingdom}
\author{Alessandro De Vita}
\affiliation{Department of Physics, King's College London, Strand, London
	WC2R 2LS, United Kingdom} 
\affiliation{Dipartimento di Ingegneria e
	Architettura, Università di Trieste, via A. Valerio 2, I-34127 Trieste, Italy}

\begin{abstract}
We provide a definition and explicit expressions for $n$-body Gaussian Process (GP) kernels, which can learn any interatomic interaction occurring in a physical system, up to $n$-body contributions, for any value of $n$.   
The series is complete, as it can be shown that the \q{universal approximator} squared exponential kernel can be written as a sum of $n$-body kernels. 
These recipes enable the choice of optimally efficient force models for each target system, as confirmed by extensive testing on various materials.
We furthermore describe how the $n$-body kernels can be \q{mapped} on equivalent representations that provide database-size-independent predictions and are thus crucially more efficient. 
We explicitly carry out this mapping procedure for the first non-trivial (3-body) kernel of the series, and we show that this reproduces the GP-predicted forces with $ \text{meV/\AA}$ accuracy while being orders of magnitude faster.   
These results pave the way to using novel force models (here named \q{M-FFs}) that are computationally as fast as their corresponding standard parametrised $n$-body force fields, while retaining the nonparametric character, the ease of training and validation, and the accuracy of the best recently proposed machine learning potentials.  

\end{abstract}
\maketitle

\section{Introduction}

Since their conception, first-principles molecular dynamics (MD) simulations \cite{Car:1985ix} based on density functional theory (DFT) \cite{Hohenberg:1964vt,Kohn:1965js} have proven extremely useful to investigate complex physical processes that require quantum accuracy. 
These simulations are computationally expensive, and thus still typically limited to hundreds of atoms and the picosecond timescale. 
For larger systems that are non-uniform and thus intractable using periodic boundary conditions, multiscale embedding (\q{QM/MM}) approaches can sometimes be used successfully.  
This is possible if full quantum accuracy is only needed in a limited (QM) zone of the system, while a simpler molecular mechanics (MM) description suffices everywhere else.
Very often, however, target problems require minimal model system sizes and simulation times so large that the calculations must be exclusively based on classical force fields i.e., force models that use the position of atoms as the only explicitly represented degrees of freedom. 

In the remainder of this introductory section we briefly review the relative strengths and weaknesses of standard parametrized (P-FFs) and machine learning force fields (ML-FFs). 
We then consider how accurate P-FFs are hard to develop but eventually fully exploit useful knowledge on the systems, while GP-based ML-FFs offer a general mathematical framework for handling training and validation, but are significantly slower (Section I A).  
These shortcomings motivate an analysis of how prior knowledge such as symmetry has been so far incorporated in GP kernels (Section I B)  and points to features still missing in ML kernels, which are commonplace in the more standard, highly efficient P-FFs based on truncated $n$-body expansions (Section I C).   
This suggests the possibility of defining a series of  $n$-body GP kernels (Section II B), providing a scheme to construct them (Section II C and D) and, after the best value of $n$ for the given target system has been identified with appropriate testing (Section II E), exploiting their dimensionally-reduced feature spaces to massively boost the execution speed of force prediction (\mbox{Section III}).

\subsection{Parametrized and machine learning force fields}

Producing accurate and fully transferable force fields is a remarkably difficult task. 
The traditional way to do this involves adjusting the parameters of carefully chosen analytic functions in the hope of matching extended reference data sets obtained from experiments or quantum calculations  \cite{ Stillinger:1985zz,Tersoff:1988jt}.
The descriptive restrictiveness of the parametric functions used is both a drawback and a strength of this methodology. 
The main difficulty is that developing a good parametric function requires a great deal of chemical intuition and patient effort, guided by trial and error steps with no guarantee of success \cite{Brenner:2000uh}.   
However, for systems and processes in which the approach is fruitful, the development effort is amply rewarded by the opportunity to provide extremely fast and accurate force models \cite{Mishin:2004bh, vanDuin:2001ig,Cisneros:2016hi,Reddy:2016ic}.
The identified functional forms will in these cases contain valuable knowledge on the target system, encoded in a compact formulation that still accurately captures the relevant physics. 
Such knowledge is furthermore often transferable to novel (while similar) systems as a \q{prior} piece of information, i.e., it constitutes a good working hypothesis on how these systems will behave. 
When QM data on the novel system become available, this can be simply used to fine-tune the parameters of the functional form to a new set of best-fit values that maximise prediction accuracy. 

Following a different approach, \q{nonparametric} ML force fields can be constructed, whose dependence on the atomic position is not constrained to a particular analytic form.
An implementation and tests exploring the feasibility of ML to describe atomic interactions can be found, e.g., in pioneering work by Skinner and Broughton \cite{Skinner:1995ea} that proposed using ML models to reproduce first-principles potential energy surfaces. 
More recent works implementing this general idea have been based on Neural Networks \cite{Behler:2007fe}, Gaussian Process (GP) regression \cite{Bartok:2010fj} or linear regression on properly defined bases \cite{Shapeev:2016kn}.
Current work aims at making these learning algorithms both faster and more accurate \cite{Li:2015eb,Glielmo:2017dj,Botu:2015kb, Ferre:2016uc,Podryabinkin:2017jp,Takahashi:2017th}.

As processing power and data communication bandwidth increase, and the cost of data storage decreases, modeling based on ML and direct inference promises to become an increasingly attractive option, compared with more traditional classical force field approaches. 
However, although ML schemes are general and have been shown to be remarkably accurate interpolators in specific systems, so far they have not become as widespread as it might have been expected. 
This is mainly because \q{standard} classical potentials are still orders of magnitude faster than their ML counterpart \cite{Boes:2016im}. 
Moreover, ML-FFs also involve a more complex mathematical and algorithmic machinery than the traditional compact functional forms of P-FFs, whose arguments are physically descriptive features that remain easier to visualize and interpret. 

\subsection{Prior knowledge and GP kernels}

These shortcomings provide motivation for the present work.  
The high computational cost of many ML models is a direct consequence of the general inverse relation between the sought flexibility and the measured speed of any algorithm capable of learning. 
Highly flexible ML algorithms by definition assume very little or no prior knowledge of the target systems. 
In a Bayesian context, this involves using a general prior kernel, typically aspiring to preserve the full universal approximator properties of e.g., the square exponential kernel \cite{Williams:2006vz,Bishop:998831}.
The price of such a kernel choice is that the ML algorithm will require large training databases \cite{Kearns:1994wq}, slowing down computations as the prediction time grows linearly with the database size. 

Large database sizes are not, however, unavoidable, and any data-intensive and fully flexible scheme to potential energy fitting is suboptimal by definition, as it exploits no prior knowledge of the system. 
This completely \q{agnostic} approach is at odds with the general lesson from classical potential development, indicating that it is essential for efficiency to incorporate in the force prediction model as much prior knowledge of the target system as can be made available. 
In this respect, GP kernels {\em can} be tailored to bring some form of prior knowledge to the algorithm.

For example, it is possible to include any symmetry information of the system.  
This can be done by using descriptors that are independent of rotations, translations and permutations \cite{Li:2015eb,Rupp:2015ep,Deringer:2017ea,Faber:2018fv}. 
Alternatively, one  can construct scalar-valued GP kernels that are made invariant under rotation (see e.g., \cite{Bartok:2013cs,Glielmo:2017dj}) or matrix-valued GP kernels made covariant under rotation (\cite{Glielmo:2017dj}, an idea that can be extended to higher-order tensors \cite{Bereau:2017vq,Grisafi:2017tw}).  
Invariance or covariance are in these cases obtained starting from a non-invariant representation by appropriate integration over the $SO(3)$ rotation group \cite{Glielmo:2017dj, Bartok:2013cs}.  

Symmetry aside, progress can be made by attempting to use kernels based on simple, descriptive features corresponding to low-dimensional feature spaces. 
Taking inspiration from parametrized force fields, these descriptors could e.g., be chosen to be interatomic distances taken singularly or in triplets, yielding kernels based on 2- or 3-body interactions \cite{Glielmo:2017dj, Szlachta:2014jh,Huo:2017ta}. 
Since low-dimensional feature spaces allow efficient learning (convergence is reached using small databases), to the extent that simple descriptors capture the correct physics, the GP process will be a relatively fast, while still very accurate, interpolator. 

\subsection {Scope of the present work} 

There are, however, two important aspects that have not as yet been fully explored while trying to develop efficient kernels based on dimensionally reduced feature spaces. 
Both aspects will be addressed in the present work. 

First, a systematic classification of rotationally invariant (or covariant, if matrix valued) kernels, representative of the feature spaces corresponding to $n$-body interactions is to date still missing. 
Namely, no definition or general recipe has been proposed for constructing $n$-body kernels, or for identifying the actual value (or effective interval of values) of $n$ associated with already available kernels.  
This would be clearly useful, however, as the discussion above strongly suggests that the kernel corresponding to the lowest value of $n$ compatible with the physics of the target system will be the most informationally efficient one for carrying out predictions: striking the right balance between speed and accuracy. 

Second, for any ML approach based on a GP kernel and a fixed database, the GP predictions for any target configuration are also fixed once and for all. 
For an $n$-body kernel, these predictions do not need, however, to be explicitly carried out as sums over the training dataset, as they could be approximated with arbitrary precision by \q{mapping} the GP prediction on a new representation based on the underlying $n$-body feature space. 
We note that this approximation step would make the final prediction algorithm independent of the database size, and thus in principle as fast as any classical $n$-body potential based on functional forms, while still parameter free. 
The remainder of this work explores these two issues, and it is structured as follows.

In the next Section II, after introducing the terminology and the notation (II A), we provide a definition of an $n$-body kernel (II B) and we propose a systematic way of constructing $n$-body kernels of any order $n$, showing how previously proposed approaches can be reinterpreted within this scheme (II C and D). We furthermore show, by extensive testing on a range realistic materials, how the optimal interaction order can be chosen as the lowest $n$ compatible with the required accuracy and the available computational power (II E). 
In the following Section III we describe how the predictions of \emph{n}-body GP kernels can be recast (mapped) with arbitrary accuracy into very fast nonparametric force fields based on machine learning (\mbox{M-FFs}) which fully retain the \emph{n}-body character of the GP process from which they were derived. 
The procedure is carried out explicitly for a 3-body kernel, and we find that evaluating atomic forces is orders of magnitude faster than the corresponding GP calculation. 

\section{$\boldsymbol{n}$-body expansions with $\boldsymbol{n}$-body kernels }

\subsection{Notation and terminology}

GP-based potentials are usually constructed by assigning an energy $\varepsilon$ to a given atomic configuration $\rho$, typically including a central atom and all its neighbors up to a suitable cutoff radius.  
The existence of a corresponding local energy function $\varepsilon(\rho)$ is generally assumed, in order to provide a total energy expression and guarantee a linear scaling of the predictions with the total number of atoms in the system. 
Within GP regression this function is calculated from a database $\mathcal{D}=\{\rho_{d},\varepsilon_{d},\mathbf{f}_{d}\}_{d=1}^{N}$
of reference data, typically obtained by quantum mechanical simulations, and usually consisting of a set of $N$ atomic configurations $\{\rho_{d}\}$ together with their relative energies $\{\varepsilon_{d}\}$ and/or
forces $\{\mathbf{f}_{d}\}$. 

It is worth noting here that although there is no well defined local
atomic energy in a reference quantum simulation, one can always use gradient
information (atomic forces, which are well defined local physical quantities) to machine-learn a potential energy function.
This can be done straightforwardly using derivative kernels
(cf., e.g., Ref. \cite{Williams:2006vz} or Ref. \cite{Macedo:2008vq}) to learn and predict forces.  
Alternatively, one can learn forces directly without an intermediate energy expression, as done in Refs. \cite{Li:2015eb,Botu:2014kc} or more recently in Ref. \cite{Glielmo:2017dj}.
A necessary condition for any of these approaches to produce energy-conserving force fields (i.e., fields that make zero work on any closed trajectory loop) is that the database is constructed once and for all, and never successively updated.   
After training on the given fixed database, the GP prediction on a target configuration $\rho$ consists of a linear combination of the kernel function values measuring the similarity of the target configuration with each
database entry: 
\begin{equation}
\varepsilon(\rho)=\sum_{d=1}^{N}k(\rho,\rho_{d})\alpha_{d},\label{eq: GP_pred}
\end{equation}
where the coefficients $\alpha_{d}$ are obtained by means of inversion of the \emph{covariance matrix} \cite{Williams:2006vz} and can be shown to minimise the regularised quadratic error between GP predictions and reference calculations.

\subsection{Definition of an $\boldsymbol{n}$-body kernel}

Classical interatomic potentials are often characterized by the number
of atoms (\q{bodies}) they let interact simultaneously (cf. e.g., Refs. \cite{Reddy:2016ic,Cisneros:2016hi}). 
To translate this concept into the realm of GP regression, we assume that the target configuration $\rho(\{\mathbf{r}_{i} \})$ represents the local atomic environment of an atom fixed at the origin of a Cartesian reference frame, expressed in terms of the relative positions $\mathbf{r}_{i}$ of the surrounding atoms. 
We define the order of a kernel $k_{n}(\rho,\rho')$
as the smallest integer $n$ for which the following property holds true: 
\begin{equation}
\frac{\partial^{n}k_{n}(\rho,\rho')}{\partial\mathbf{r}_{i_1}\cdots\partial\mathbf{r}_{i_n}}=0\label{eq: nBody_ker_def}
\hspace{1em} \hspace{1em}
\forall\mathbf{r}_{i_1}\neq\mathbf{r}_{i_2}\neq\dots\neq\mathbf{r}_{i_n},
\end{equation}
where $\mathbf{r}_{i_1},\dots,\mathbf{r}_{i_n}$ are the positions of any choice of a set of $n$ different surrounding atoms. 
By virtue of linearity, the local energy in Eq.~(\ref{eq: GP_pred}) will also satisfy the same property if $k_{n}$ does.  
Thus, Eq.~(\ref{eq: nBody_ker_def}) implies that the central atom in a local configuration interacts with up to $n-1$ other atoms simultaneously, making the interaction energy term $n$-body. 
For instance, using a 2-body kernel, the force on the central atom due to atom $\mathbf{r}_{j}$ will not depend on the position of any other atom $\mathbf{r}_{l\neq j}$ belonging to the target configuration $\rho(\{\mathbf{r}_{i} \})$. 
Eq.~(\ref{eq: nBody_ker_def}) can be used directly to check through either numeric or symbolic differentiation if a given kernel is of order $n$, a fact that might be far from obvious from its analytic form, depending on how the kernel is built. 

\subsection{Building $\boldsymbol{n}$-body kernels I: SO(3) integration }

Following a standard route \cite{Ferre:2015dq,Bartok:2013cs,Glielmo:2017dj},
we begin by representing each local atomic configuration as a sum of Gaussian functions $\mathcal{N}$ with a given variance $\sigma^{2}$, centered on the  $M$ atoms of the configuration: 
\begin{equation}
\rho(\mathbf{r}, \{ \mathbf{r}_i  \} )=\sum_{i=1}^{M}\mathcal{N}(\mathbf{r}\mid\mathbf{r}_{i},\sigma^{2}),\label{eq: Gauss_exp}
\end{equation}
where $\mathbf{r}$ and $\{\mathbf{r}_{i}\}_{i=1}^M$ are position vectors relative to the central atom of the configuration. 
This representation guarantees by construction invariance with respect to translations and permutations of atoms (here assumed to be of a single chemical species). 
As described in \cite{Glielmo:2017dj}, a covariant 2-body 
force kernel can be constructed from the non-invariant scalar (\q{base})  
kernel obtained as a dot product overlap integral of the two configurations:  
\begin{align}
k_{2}(\rho,\rho') & =\int d\mathbf{r}\,\rho(\mathbf{r})\rho'(\mathbf{r}) \nonumber \\
 & =L\sum_{\substack{i\in\rho, 
j\in\rho'
}
}\mathrm{e}^{-(\mathbf{r}_{i}-\mathbf{r}_{j}')^{2}/4\sigma^{2}},\label{eq: k2}
\end{align}
where $L$ is an unessential constant factor, omitted for convenience from now on. 
That (\ref{eq: k2}) is a 2-body kernel consistent with the definition of Eq.~(\ref{eq: nBody_ker_def}) can be checked
straightforwardly by explicit differentiation (see Appendix A).
Its $2$-body structure is also readable from the fact that $k_{2}$ is a sum of contributions comparing pairs of atoms in the two configurations, the first pair located at the two ends of vector $\mathbf{r}_{i}$ in the target configuration $\rho$, and consisting of the central atom and atom $i$, 
and the second pair similarly represented by the vector $\mathbf{r}_{j}^{\prime}$ in the database configuration $\rho'$. 
A rotation-covariant matrix-valued force kernel can at this point be constructed by Haar integration \cite{Nachbin:1965ti,SchulzMirbach:1994jn} as an integral  over the $SO(3)$ manifold \cite{Glielmo:2017dj}: 
\begin{align}
\mathbf{K}_{2}^{s}(\rho,\rho') & =\int_{SO(3)} d\mathcal{R} \, \mathbf{R} \, k_{2}(\rho, \mathcal{R}\rho' ). \label{eq: k_cov}
\end{align}
This kernel can be used to infer forces on atoms using a GP regression vector formula analogous to Eq.~\eqref{eq: GP_pred} (see Ref. \cite{Glielmo:2017dj}).  
These forces belong to a 2-body force field purely as a consequence of the base kernel property in Eq.~(\ref{eq: nBody_ker_def}). 
It is interesting to notice that there is no use or need for an intermediate energy expression to construct this 2-body force field, which is automatically energy-conserving.
Higher order $n$-body base kernels can be constructed as finite powers of the 2-body base kernel (\ref{eq: k2}): 
\begin{equation}
k_{n}(\rho,\rho')=k_{2}(\rho,\rho')^{n-1}\label{eq: kn}
\end{equation}
where the $n$-body property (Eq.~\eqref{eq: nBody_ker_def})
can once more be checked by explicit differentiation (see Appendix A). 
Furthermore, taking the exponential of the kernel in Eq.~(\ref{eq: k2}) gives rise to a fully many-body base kernel, as all powers of $k_{2}$ are contained in the exponential formal series expansion: 
\begin{align}
k_{MB}(\rho,\rho') & = \mathrm{e}^{k_{2}(\rho,\rho')/\theta^{2}}\nonumber \\
 & =1+\frac{1}{\theta^{2}}k_{2}+\frac{1}{2!\theta^{4}}k_{3}+\dots.\label{eq: k_SE}
\end{align}
One can furthermore check that the simple exponential many-body kernel $k_{MB}$ defined above is, up to normalisation, equivalent to the squared exponential kernel \cite{Williams:2006vz} on the natural distance induced by the dot product kernel $k_{2}(\rho,\rho')$: 
\begin{align}
\mathrm{e}^{-d^{2}(\rho,\rho')/2\theta^{2}} = & N(\rho)N(\rho')k_{MB}(\rho,\rho')\label{eq: Sq_exp}  \\
d^{2}(\rho,\rho') = & k_{2}(\rho,\rho)+k_{2}(\rho',\rho')-2k_{2}(\rho,\rho').\label{eq: distance}
\end{align}

\begin{figure}
	\includegraphics[width=1.0\columnwidth]{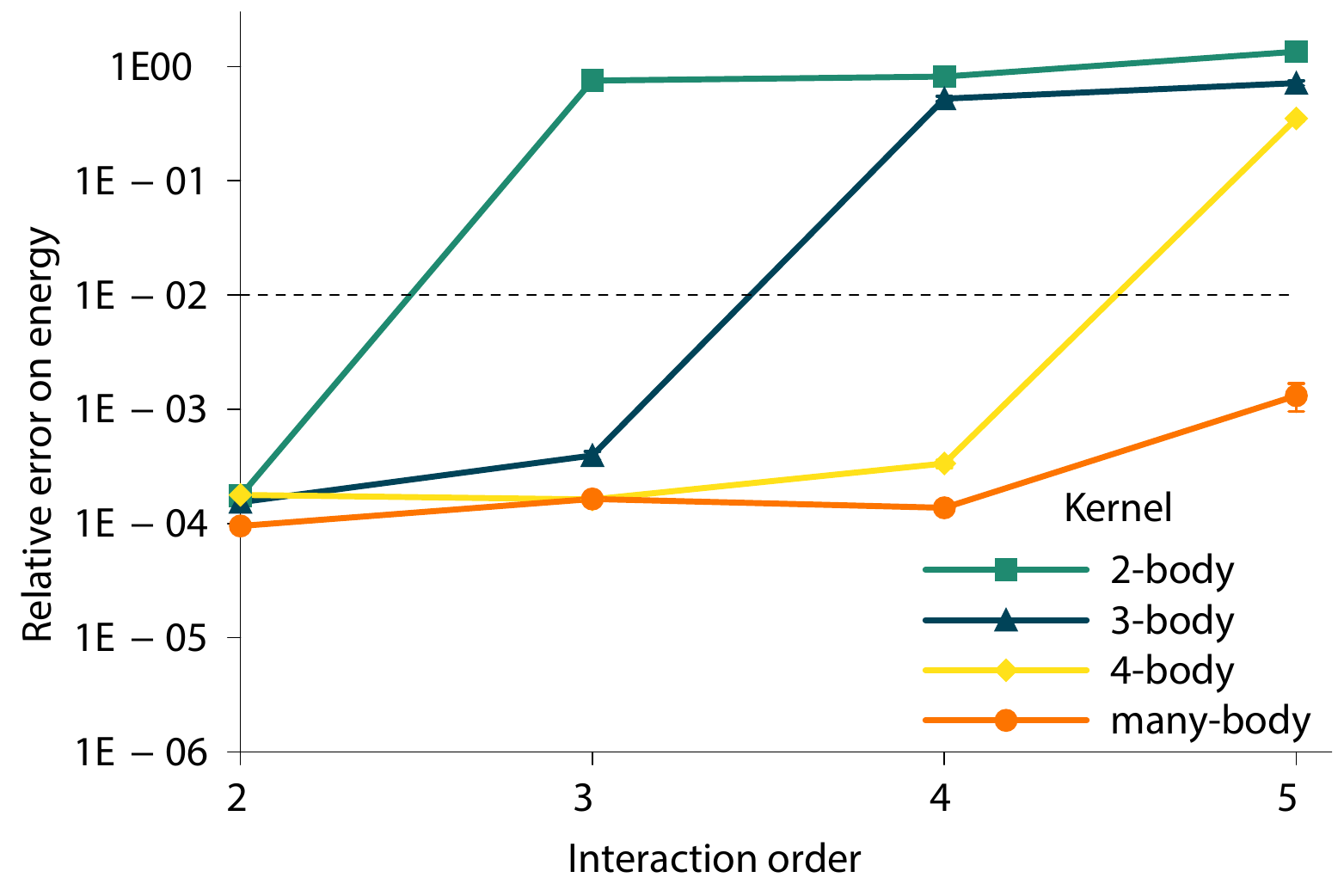}
	\caption{\textcolor{black}{GP relative error as a function of the interaction order (2- to 5-body), using $n$-body kernels with increasing $n$. Learning energies within baseline precision (black dashed line) requires an $n$-body kernel with $n$ at least as high as the particles' interaction order.} \label{fig: simple_test}}
\end{figure}

To check on these ideas, we next test the accuracy of these kernels in learning the interactions occurring in a simple 1D model consisting of $n'$ particles interacting via an \emph{ad-hoc} $n'$-body potential (see Appendix B.).
We first let the particles interact to generate a configuration database, and then attempt to machine-learn these interactions using the kernels just described.
Figure \ref{fig: simple_test} illustrates the average prediction errors
on the local energies of this system incurred by the GP regression based on four different kernels as a function of the interaction order $n'$.  
It is clear from the graph that a force field that lets the $n'$ particles interact simultaneously can only be learned accurately with a ($n\geq n'$)-body kernel (\ref{eq: kn}), or with the many-body exponential kernel
(\ref{eq: k_SE}) which contains all interaction orders. 

To construct $n$-body kernels useful for applications to real 3D systems we need to include rotational symmetry by averaging over the rotation group. 
For our present scopes, it is sufficient to discuss the case of rotation-invariant $n$-body scalar energy kernels, for which the integral (formally a \emph{transformation integration} \cite{Haasdonk:2007ff}) is readily obtained from Eq.~(\ref{eq: k_cov}) by simply dropping the $\mathbf{R}$ matrix in the integrand: 
\begin{align}
k_{n}^{s}(\rho,\rho') & =\int_{SO(3)}d\mathcal{R}\:k_{n}(\rho, \mathcal{R}\rho').\label{eq: SO3_int}
\end{align}
The use of this integral in the context of potential energy learning was originally proposed in \cite{Bartok:2013cs}, where it was carried out using appropriate functional expansions.
Alternatively, one can exploit the Gaussian nature of the configuration expansion (\ref{eq: Gauss_exp}) to obtain an analytically exact formula, as done further below.
The resulting symmetrized $n$-body kernel $k_{n}^{s}$ will learn faster than its non-symmetrized counterpart $k_{n}$, as the rotational degrees of freedom
have been integrated out. 
This is because a non-symmetrized $n$-body kernel ($k_{n}$) must learn functions of $3n-3$ variables (translations are taken into account by the local  representation based on relative position in Eq.~\eqref{eq: Gauss_exp}).
After integration, the new kernel $k_{n}^{s}$ defines a smaller and more physically-based space of functions of $3n-6$ variables, which is the rotation-invariant functional domain of $n$ interacting particles. 

The symmetrization integral in Eq. (\ref{eq: SO3_int}) can be written down for the many-body base kernel $k_{MB}$ (Eq.~(\ref{eq: k_SE})), to define a new 
many-body kernel $k^{s}_{MB}$ invariant under all physical symmetries: 
\begin{align}
k_{MB}^{s}(\rho,\rho') & =\int_{SO(3)}d\mathcal{R}\:k_{MB}(\rho, \mathcal{R}\rho').\label{eq: SE_SO3_int}
\end{align}
By virtue of the universal approximation theorem \cite{Hornik:1993dh,Williams:2006vz} 
this kernel would be able to learn arbitrary physical interactions with arbitrary accuracy, if provided with sufficient data.  

Unfortunately, the exponential kernel (\ref{eq: k_SE}) has to date resisted all attempts to carry out the analytic integration over rotations 
\eqref{eq: SE_SO3_int}, leaving as the only open options numerical integration, or discrete summation over a relevant point group of the system \cite{Glielmo:2017dj}. 
On the other hand, the analytic integration of 2- or 3-body kernels to give symmetric \textit{n}-body kernels can be carried out in different ways.	

For example, one could use an intermediate step that was introduced during the construction of the widely used SOAP kernel  \cite{Bartok:2013cs,Szlachta:2014jh,Thompson:2015dw,De:2016ia,Rowe:2017tw}.  
This kernel has a full many-body character \cite{Bartok:2013cs}, ensured by the prescribed normalisation step defined by Eqs. (31-36) of the standard Ref. \cite{Bartok:2013cs}), which made it possible to use it e.g., to augment to full many-body the descriptive power of a 2- and 3-body explicit kernel expansion \cite{Deringer:2017ea}.  
However, the Haar integral over rotations introduced in \cite{Bartok:2013cs} as an intermediate kernel construction step could also be seen, if taken on its own, as a transformation integration procedure \cite{Haasdonk:2007ff} yielding a symmetrized \textit{n}-body kernel as defined in Eq.~\eqref{eq: SO3_int} above, which would in turn become a higher finite-order kernel if raised to integer powers $\zeta \geq 2 $ (see next subsection).  

Carrying out Haar integrals is not, in general, an easy task. 
In the example above, computing a general rotation invariant $n$-body kernel via the exact, suitably truncated spherical harmonics expansion procedure of Ref. \cite{Bartok:2013cs} becomes challenging for $n>3$.  
Significant difficulties likewise arise if attempting a \q{covariant} integration over rotations, for which we found an exact analytic expression only for 2- and 3-body matrix-valued kernels \cite{Glielmo:2017dj}, with a technique that becomes unviable for $n>3$. 
Fortunately, the Haar integration can be avoided altogether, following the simple route of constructing symmetric \textit{n}-kernels directly using symmetry-invariant descriptors, as we will see in the next section.  
The problem of obtaining an analytic Haar integral expression for the general \textit{n} case remains, however, an interesting one, which we tackle in the remainder of this section following a novel analytic route. 

\begin{figure}
	\includegraphics[width=1.0\columnwidth]{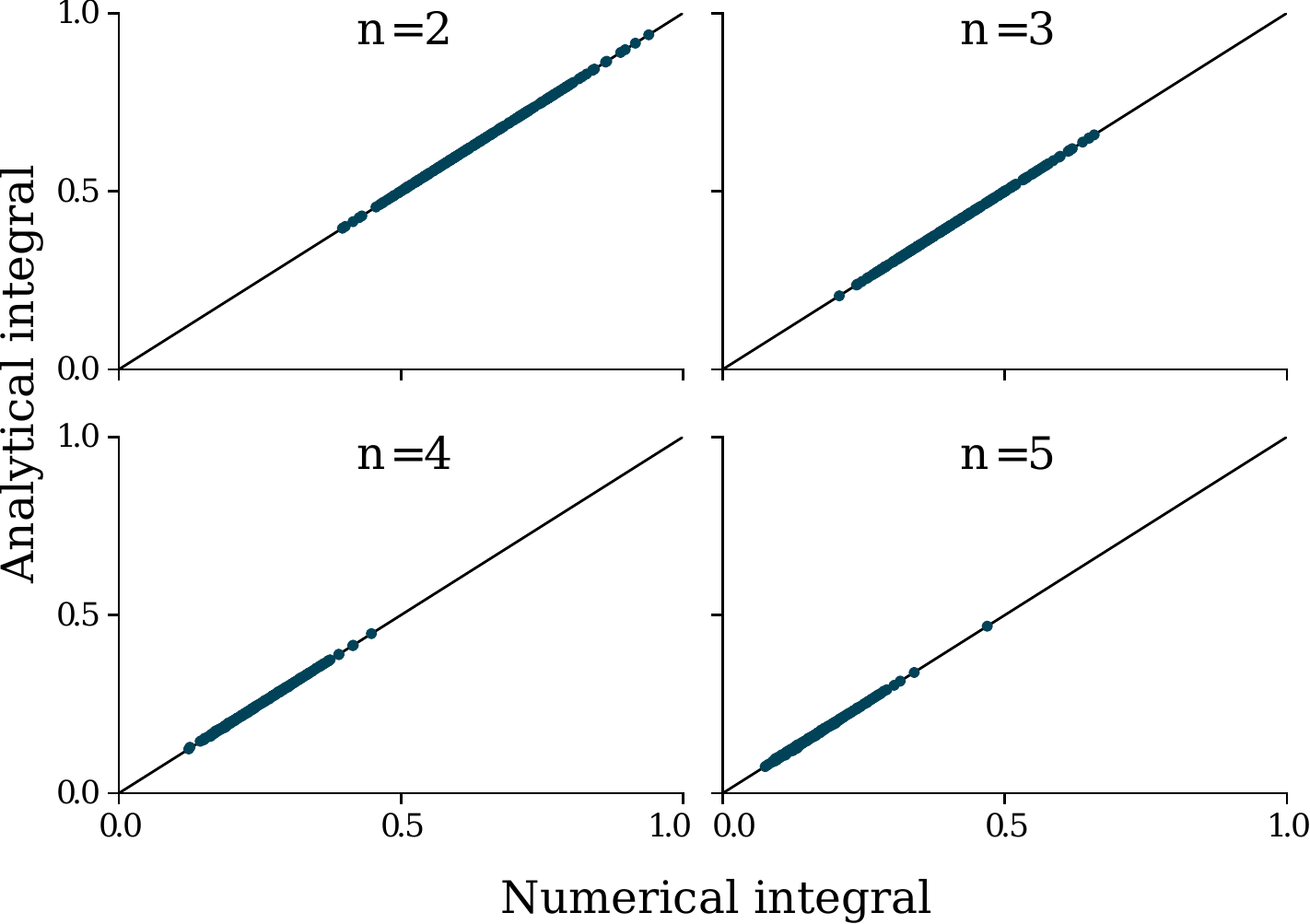}
	\caption{\textcolor{black}{Scatter plots showing the values of the integral in \eqref{eq: kn_decomposition_2} (on a random selection of configurations) computed either by numerical integration or via the analytic expression (Eqs. (\ref{eq: k_tilde_int_eq_CI}, \ref{eq: C_mathcal}, \ref{eq: JAMES})). Interaction orders from $n=2$ to $n=5$ are considered.} \label{fig: num_vs_an}}
\end{figure}

We first write the $n$-body base kernel of Eq.~\eqref{eq: kn} as an explicit product of $(n-1)$ 2-body kernels. 
The Haar integral \eqref{eq: SO3_int} can then be written as
\begin{align}\label{eq: kn_decomposition}
&k_{n}^{s}(\rho,\rho') =\!\!\!\!\!\! \sum_{\substack{\mathbf{i}=(i_{1},\dots,i_{n-1})\in\rho\\
		\mathbf{j}=(j_{1},\dots,j_{n-1})\in\rho'}} \!\!\!\!\!\!
	\tilde{k}_{\mathbf{i}, \mathbf{j}} \\ 
	\tilde{k}_{\mathbf{i}, \mathbf{j}}  =& \int d\mathcal{R} \, \mathrm{e}^{-\frac{\|\mathbf{r}_{i_1}-\mathbf{R}\mathbf{r}_{j_1}'\|^{2}}{4\sigma^{2}}} \dots \mathrm{e}^{-\frac{\|\mathbf{r}_{i_{n-1}}-\mathbf{R}\mathbf{r}_{j_{n-1}}'\|^{2}}{4\sigma^{2}}}
	\label{eq: kn_decomposition_2}
\end{align}
where now for each of the two configurations $\rho,\rho'$, the sum runs  
over all $n$-plets of atoms that include the central atom (whose indices $i_0$ and $j_0$ are thus omitted). 
Expanding the exponents as $(\mathbf{r}_{i}-\mathbf{R}\mathbf{r}_{j}')^2 = r_{i}^2  + r_{j}^{\prime 2} - 2 \text{\rm{Tr}} ( \mathbf{R} \mathbf{r}_{j}' \mathbf{r}_{i}^{\rm{T}} )$
allows us to extract from the integral \eqref{eq: kn_decomposition_2}
a rotation independent constant $\mathcal{C}_{\mathbf{i}, \mathbf{j}}$, 
and to express the rotation-dependent scalar products sum 
as a trace of a matrix product:    
\begin{align}
\tilde{k}_{\mathbf{i}, \mathbf{j}} & = \mathcal{C}_{\mathbf{i}, \mathbf{j}} \mathcal{I}_{\mathbf{i}, \mathbf{j}} \label{eq: k_tilde_int_eq_CI}
	\\
	\mathcal{C}_{\mathbf{i}, \mathbf{j}} & = \mathrm{e}^{-(r_{i_1}^2  + r_{j_1}^{\prime 2} + \dots r_{i_{n-1}}^2 + r_{j_{n-1}}^{\prime 2})/4 \sigma^2} \label{eq: C_mathcal}
	\\
	\mathcal{I}_{\mathbf{i}, \mathbf{j}} & = \int d\mathcal{R} \, \mathrm{e}^{\rm{Tr}(\mathbf{R} \mathbf{M}_{\mathbf{i}, \mathbf{j}})} \label{eq: I_mathcal}
\end{align}
where the matrix $\mathbf{M}_{\mathbf{i}, \mathbf{j}}$ is the sum of the outer products of the ordered vector couples in the two configurations:  $\mathbf{M}_{\mathbf{i}, \mathbf{j}} = (\mathbf{r}_{j_1}' \mathbf{r}_{i_1}^{\rm{T}} + \dots + \mathbf{r}_{j_{n-1}}' \mathbf{r}_{i_{n-1}}^{\rm{T}})/2\sigma^2$.
The integral \eqref{eq: I_mathcal} occurs in the context of multivariate statistics as the generating function of the non-central Wishart distribution \cite{Anderson:1946ju}. 
As shown in \cite{James:1955bo}, it can be expressed as a power series in the symmetric polynomials ($\alpha_1 = \sum_i^3 \mu_i, \alpha_2 = \sum_{i<j}^3 \mu_i \mu_j, \alpha_3 = \mu_1\mu_2\mu_3$) of the eigenvalues $\{ \mu_i \}_{i=1}^3$ of the symmetric matrix $\mathbf{M}_{\mathbf{i}, \mathbf{j}}^{\rm{T}} \mathbf{M}_{\mathbf{i}, \mathbf{j}}$:
\begin{align} \label{eq: JAMES}
\mathcal{I}_{\mathbf{i}, \mathbf{j}} & = \sum_{p_1,p_2,p_3} A_{p_1 p_2 p_3} \alpha_1^{p_1}\alpha_2^{p_2}\alpha_3^{p_3}
	\\
A_{p_1 p_2 p_3}	& = \frac{ \pi \, 2^{-(1+2p_1+4p_2+6p_3)} (p_1+2p_2+4p_3)! }{p_1!p_2!p_3! \Gamma(\frac{3}{2}+p_1+2p_2+3p_3)  \Gamma(1+p_2+2p_3)} \nonumber
	\\
&\quad \times \frac{1}{\Gamma(\frac{1}{2}+p_3) (p_1+2p_2+3p_3)!} . \label{eq: JAMES_2}
\end{align}

Remarkably, in this result (whose exactness is checked numerically in Figure \ref{fig: num_vs_an}) the integral over rotations does not depend on the order $n$ of the base kernel, once the matrix $\mathbf{M}_{\mathbf{i},\mathbf{j}}$ is computed.   
This is not the case for previous approaches to integrating over rotations \cite{Bartok:2013cs,Glielmo:2017dj} that need to be reformulated with increasing and eventually prohibitive difficulty each time the order $n$ 
needs to be increased. 

However, the final expression given by Eqs. (\ref{eq: k_tilde_int_eq_CI}-\ref{eq: JAMES_2}) is still a relatively complex and computationally heavy function of the atomic positions. 
To alleviate its evaluation cost, it would be interesting to see whether it is possible to recast it as an explicit scalar product in a given feature space. 
This would allow e.g.,  to transfer most of the computational burden to the pre-computation of the corresponding basis functions. 
Fortunately such complexity can be largely avoided altogether if equally accurate kernels can be built by physical intuition at least for the most relevant lowest $n$ orders, as discussed in the next section.

\subsection{Building $\boldsymbol{n}$-body kernels II: $\boldsymbol{n}$-body feature spaces and uniqueness issues}

The practical effect of the Haar integration (\ref{eq: SO3_int}) is the elimination of the three spurious rotational degrees of freedom. 
The same result can always be achieved by selecting a group of symmetry- invariant degrees of freedom for the system, typically including the distances and/or bond angles found in local atomic environments, or simple functions of these. 
Appropriate symmetrized kernels can then simply be obtained by defining a similarity measure \emph{directly} on these invariant quantities \cite{Li:2015eb,Rupp:2012kxa,Deringer:2017ea,Faber:2018fv}.
To construct symmetry invariant $n$-body kernels with $n=2$ and $n=3$ we can choose these degrees of freedom to be just interparticle distances: 
\begin{align}
\ k_{2}^{s}(\rho,\rho') & =\sum_{\substack{i\in\rho\\
j\in\rho'
}
}\tilde{k}_{2}(r_{i},r_{j})\label{eq: k2_k3_q}\\
k_{3}^{s}(\rho,\rho') & =\sum_{\substack{i_1,i_2\in\rho\\
j_1,j_2\in\rho'
}
}\tilde{k}_{3}((r_{i_1},r_{i_2},r_{i_1 i_2}),(r_{j_1}',r_{j_2}',r_{j_1j_2}')) \label{3bodyk}
\end{align}
where the $\tilde{k}$ are kernel functions that directly specify the correlation of distances, or triplets of distances, found within the two configurations.
Since  these kernels learn functions of low-dimensional spaces, their exact analytic form is not essential for performance, as any fully non-linear function $\tilde{k}$ will give equivalent converged results in the rapidly reached large-database limit. 
\begin{figure}
	\includegraphics[width=1.0\columnwidth]{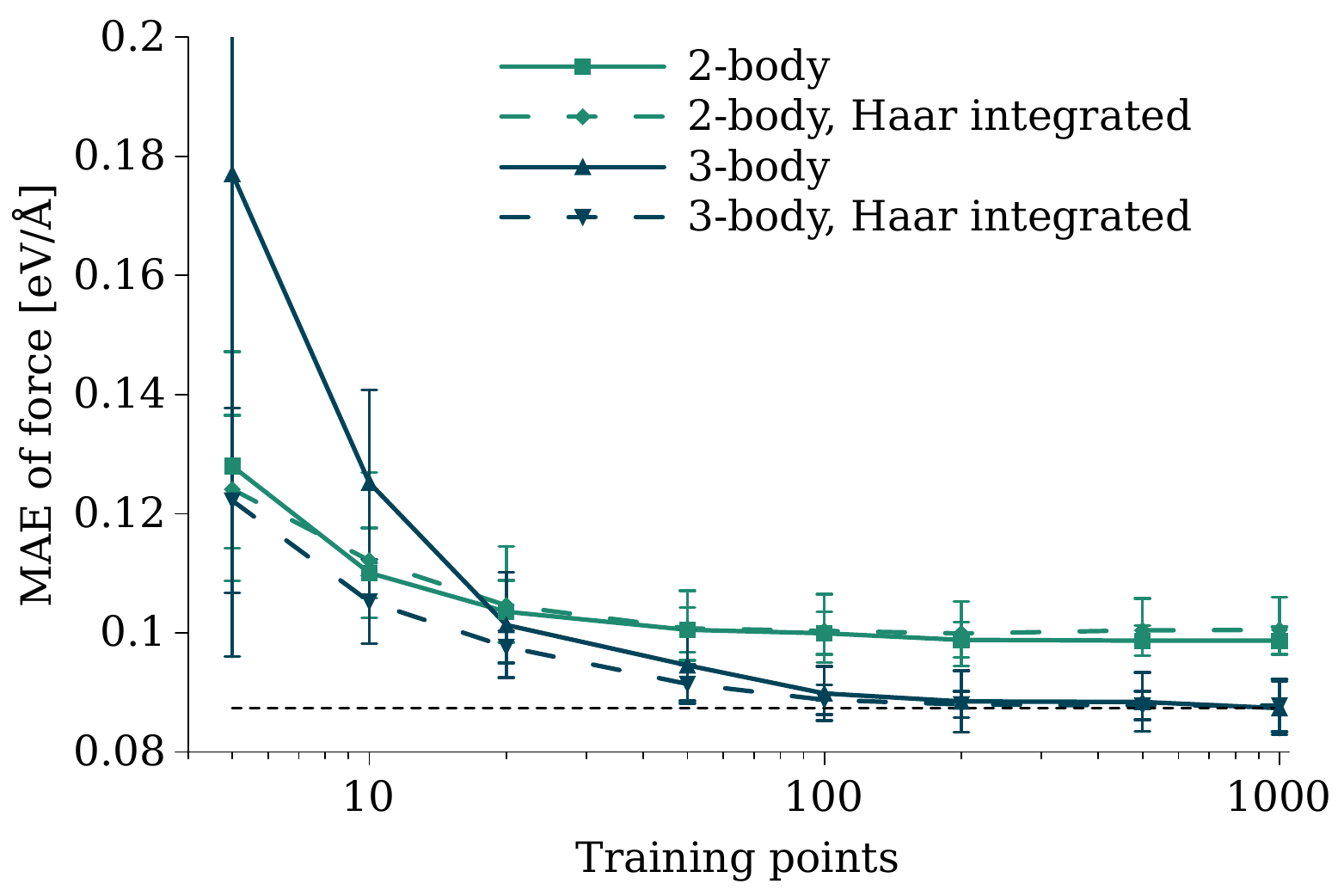}
	\caption{\textcolor{black}{Learning curves for 2 and 3-body kernels obtained either via a Haar integration (Eqs. (\ref{eq: kn_decomposition}-\ref{eq: JAMES_2})), or directly specifying a similarity kernel function of the effective degrees of freedom (Eqs. (\ref{eq: k2_k3_q}, \ref{3bodyk})).
	}}
	\label{fig: kn_and_kn_int}
\end{figure}
This equivalence can be neatly observed in Figure \ref{fig: kn_and_kn_int}, which reports the performance of 2- and 3-body kernels built either directly over the set of distances (Eqs. \eqref{eq: k2_k3_q} and  \eqref{3bodyk}) or via the exact Haar integral (Eqs. (\ref{eq: kn_decomposition}-\ref{eq: JAMES_2})). 
As the test system is crystalline Silicon, 3-body kernels are better performing. 
However, since convergence of the 2- and 3-body feature space is quickly achieved (at about $N=50$ and $N=100$ respectively), there is no significant performance difference between $SO(3)$-integrated $n$-body kernels and physically motivated ones.
Consequently, for low interaction orders, simple and computationally fast kernels like the ones in Eqs. (\ref{eq: k2_k3_q}, \ref{3bodyk}) are always preferable to more complex (and heavier) alternatives obtained via integration over rotations (e.g., the one defined by Eqs. (\ref{eq: kn_decomposition}-\ref{eq: JAMES_2}) or those found in Refs. \cite{Bartok:2013cs,Glielmo:2017dj}.

We note at this point that Eq.~(\ref{eq: k2_k3_q}) can be generalized to construct a symmetric $n$-body \mbox{kernel}
\begin{equation}
k_{n}^{s}(\rho,\rho')=\sum_{\substack{i_{1},\dots,i_{n-1}\in\rho\\
j_{1},\dots,j_{n-1}\in\rho'
}
}\,\tilde{k}_{n}(\mathbf{q}_{i_{1},\dots,i_{n-1}},\mathbf{q}_{j_{1},\dots,j_{n-1}}'),\label{eq: kn_q}
\end{equation}
where the components of the feature vectors $\mathbf{q}$ are the chosen symmetry-invariant degrees of freedom describing the $n$-plet of atoms. 

The $\mathbf{q}$ feature vectors are required to be $(3n-6)$ dimensional for all $n$, except for $n=2$, where they become scalars. 
In practice, for $n>3$ selecting a suitable set of invariant degrees of freedom is not trivial.  
For instance, for $n=4$ the set of six unordered distances between four particles does not specify their relative positions unambiguously, while for $n>4$ the number of distances associated with $n$ atoms exceeds the target feature space dimension $3n-6$.
Meanwhile, the computational cost of evaluating the full sum in Eq.~(\ref{eq: kn_q}) very quickly becomes prohibitively large as the number of elements in the sum grows exponentially with $n$. 

The order of an \textit{already} \emph{symmetric} $n$-body kernel can however be augmented with no computational overhead by generating a derived kernel through simple exponentiation to an integer power, at the cost of losing the \emph{uniqueness} \cite{Bartok:2013cs,Huang:2016gh,vonLilienfeld:2015ba} of the representation. 
\begin{figure}
	\includegraphics[width=1.0\columnwidth]{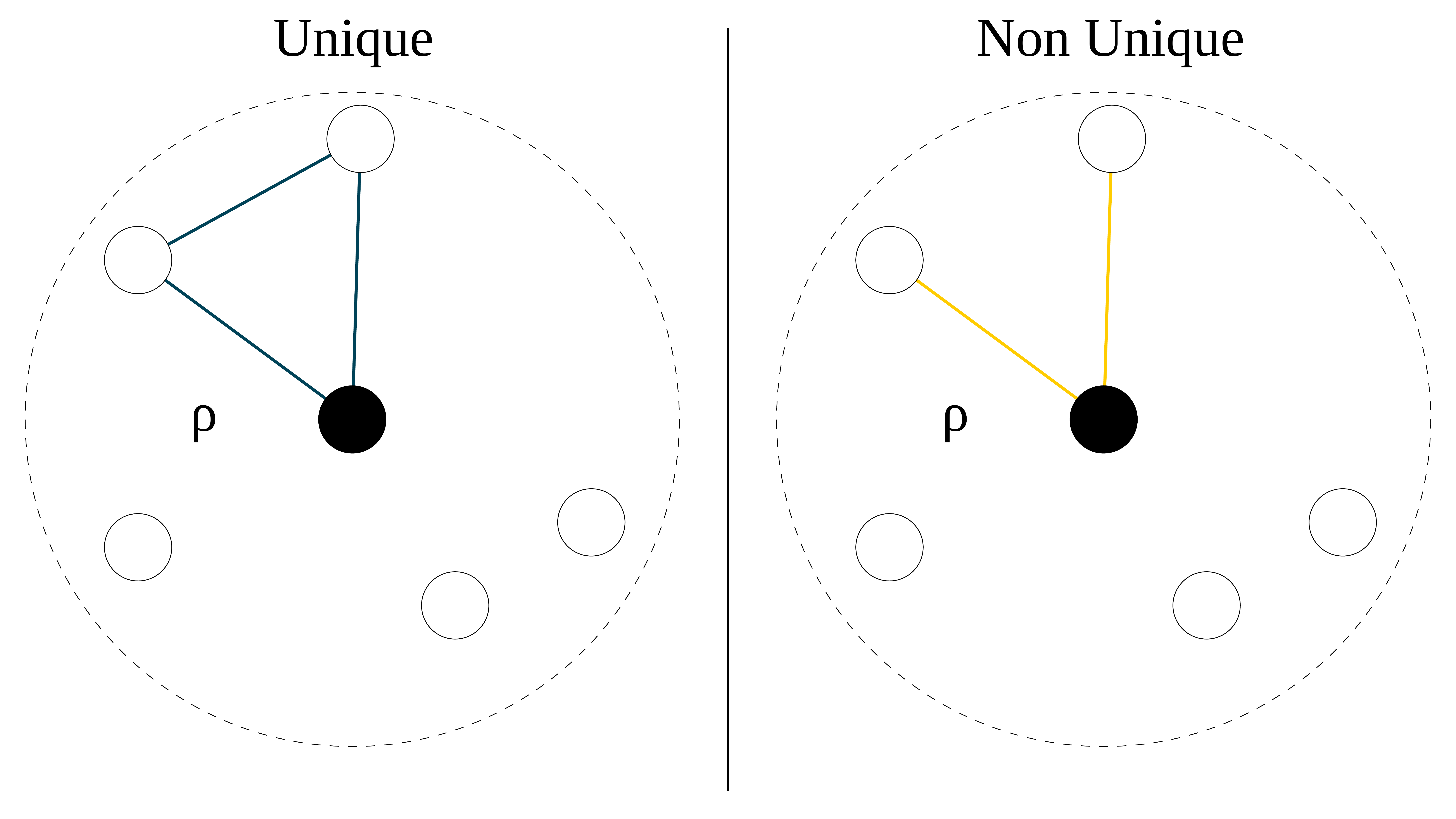}
	\caption{\textcolor{black}{Unique interaction (left panel) associated with the 3-body kernel $k_3^s$ \eqref{3bodyk} compared with the non-unique 3-body interaction (right panel) associated with the kernel $k^{\neg u}_3 = (k_2^s)^2$} \eqref{eq: non_unique_k3}, which is a function of two distances only (see text). 
		\label{fig: non_unique_3b}}
\end{figure}
This can be easily understood by means of an example (graphically illustrated in Figure \ref{fig: non_unique_3b}). 
Let us consider the 2-body symmetric kernel $k_{2}^{s}$
(Eq.~\eqref{eq: k2_k3_q}) which learns a function of just a single distance, and therefore treats the $r_{i}$ distances between the central atom and its neighbors independently. 
Its square is the kernel 
\begin{equation}
k_{3}^{\neg u}(\rho,\rho')  =\sum_{\substack{i_1,i_2\in\rho\\
		j_1,j_2\in\rho'}} \tilde{k}_{2}(r_{i_1},r_{j_1}')
        \tilde{k}_{2}(r_{i_2},r_{j_2}')
        \label{eq: non_unique_k3}
\end{equation}
which will be able to learn functions of two distances $r_{i_{1}},r_{i_{2}}$ from the central atom of the target configuration $\rho$ (see Figure \ref{fig: non_unique_3b})  and thus will be a $3$-body kernel in the sense of Eq.~(\ref{eq: nBody_ker_def}).   
However, this kernel cannot resolve angular information, as rotating the atoms in $\rho$ around the origin by independent, arbitrary angles will yield identical predictions. 

Extending this line of reasoning, it is easy to show that squaring a symmetric $3$-body kernel yields a kernel that can capture interactions up to 5-body, although again non-uniquely.     
This has often been done in practice by squaring the SOAP integral \cite{Deringer:2017ea,Rowe:2017tw}.   
In general, raising a 3-body \q{input} kernel to an arbitrary integer power $\zeta \geq 2$ yields an $n$-body output kernel of order $2\zeta+1$, $k_{n=2\zeta+1}^{\neg u}=k_{3}^{s}(\rho,\rho'){}^{\zeta}$. 
This kernel is also non-unique as it will learn a function of only $3\zeta$ variables, while the total number of relevant $n$-body degrees of freedom ($3n-6 = 6\zeta - 3$) is always larger than this. 
Substituting 3 with any $n'$ order of the symmetrized input kernel will similarly generate a $k_{n}^{\neg u}=k_{n'}^{s}(\rho,\rho'){}^{\zeta}$ kernel of order $n=(n'-1)\zeta+1$. 
A simple calculation reveals that, also in the general case, the number of variables on which $k_{n}^{\neg u}$ is implicitly built is $(3 n^{\prime}  - 6 )\zeta $, always smaller than the full dimension of $n$-body feature space $(3n' - 3)\zeta - 3$ (as expected, the two become equal only for the trivial exponent $\zeta = 1$).

None of the kernels obtained as finite powers of some symmetric lower-order kernels is a many-body one (they will all satisfy Eq.~\eqref{eq: nBody_ker_def} for some finite $n$). 
However, an attractive immediate generalization consists of substituting any squaring or cubing with full exponentiation. 
For instance, exponentiating a symmetrized 3-body kernel we obtain the many-body kernel $k_{MB}=\exp[k_{3}^{s}(\rho,\rho')]\label{eq: k_nu_SE}$.
It is clear from the infinite expansion in Eq.~(\ref{eq: k_SE}) that this kernel is a many-body one in the sense of Eq.~(\ref{eq: nBody_ker_def}), and is also fully symmetric 
\footnote{
		One could also inexpensively obtain a many body kernel by normalisation of an explicit finite order one, for instance, as $k_{MB} = k_3^s(\rho, \rho')/\sqrt{k_3^s(\rho, \rho) k_3^s(\rho', \rho')}$.
		The denominator makes this many-body in the sense of Eq. (\ref{eq: nBody_ker_def}) (as is also the case for the SOAP kernel, see discussion in Section II C, while no Haar integration is needed here). 
}. 
As is also the case for all finite-power kernels, the computational cost of this many body kernel will depend on the order $n'$ of the input kernel (3 in the present example) as the sum in Eq.~\eqref{eq: kn_q} only runs on the atomic $n'$-plets  (here, triplets) in $\rho$ and $\rho'$. 
This new kernel is not a priori known to neglect any order of interaction that might occur in a physical system and thus be encoded in a reference QM training database. 
To summarise, we provided a definition for an $n$-body kernel, and proposed a general formalism for building $n$-body kernels by exact Haar integration over rotations. 
We then defined a class of simpler kernels based on rotation invariant features that are also $n$-body according to the previous definition. 
As both approaches become computationally expensive for high values of $n$,  we pointed out that $n$-body kernels can be built as powers of lower-order input $n'$-body kernels, with no additional computational overhead.   
While such a procedure in our case comes at the cost of sacrificing the unicity property of the descriptor, it also suggests how to build, by full exponentiation, a many-body symmetric kernel.
For many applications, however, using a finite-order kernel will provide the best option. 

\begin{center}
	\begin{figure*}
		\begin{centering}
			\subfloat[\label{fig:Crystalline-Nickel}]{\includegraphics[width=0.4\paperwidth]{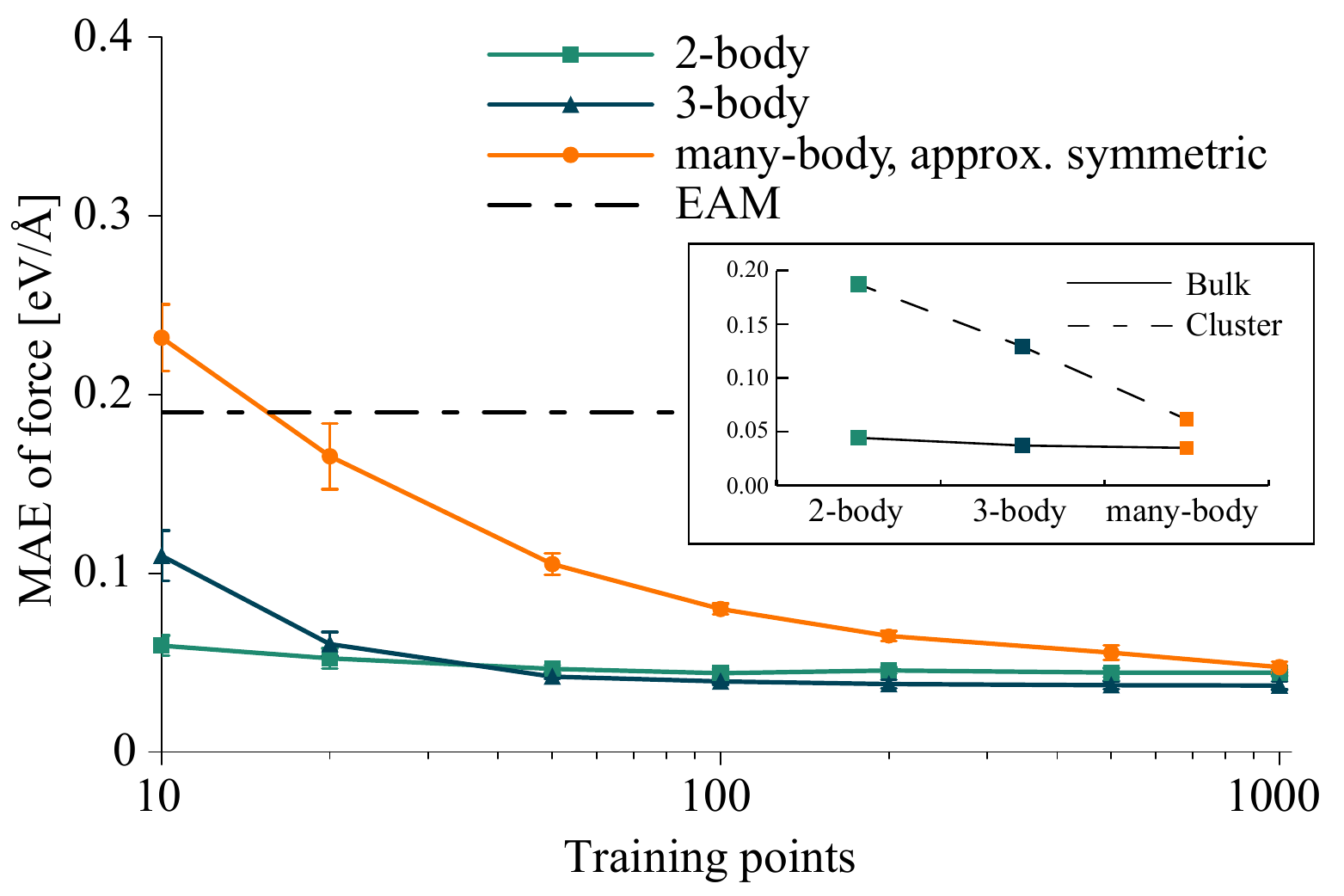}
				
			}\subfloat[\label{fig:Iron-with-vacancy}]{\includegraphics[width=0.4\paperwidth]{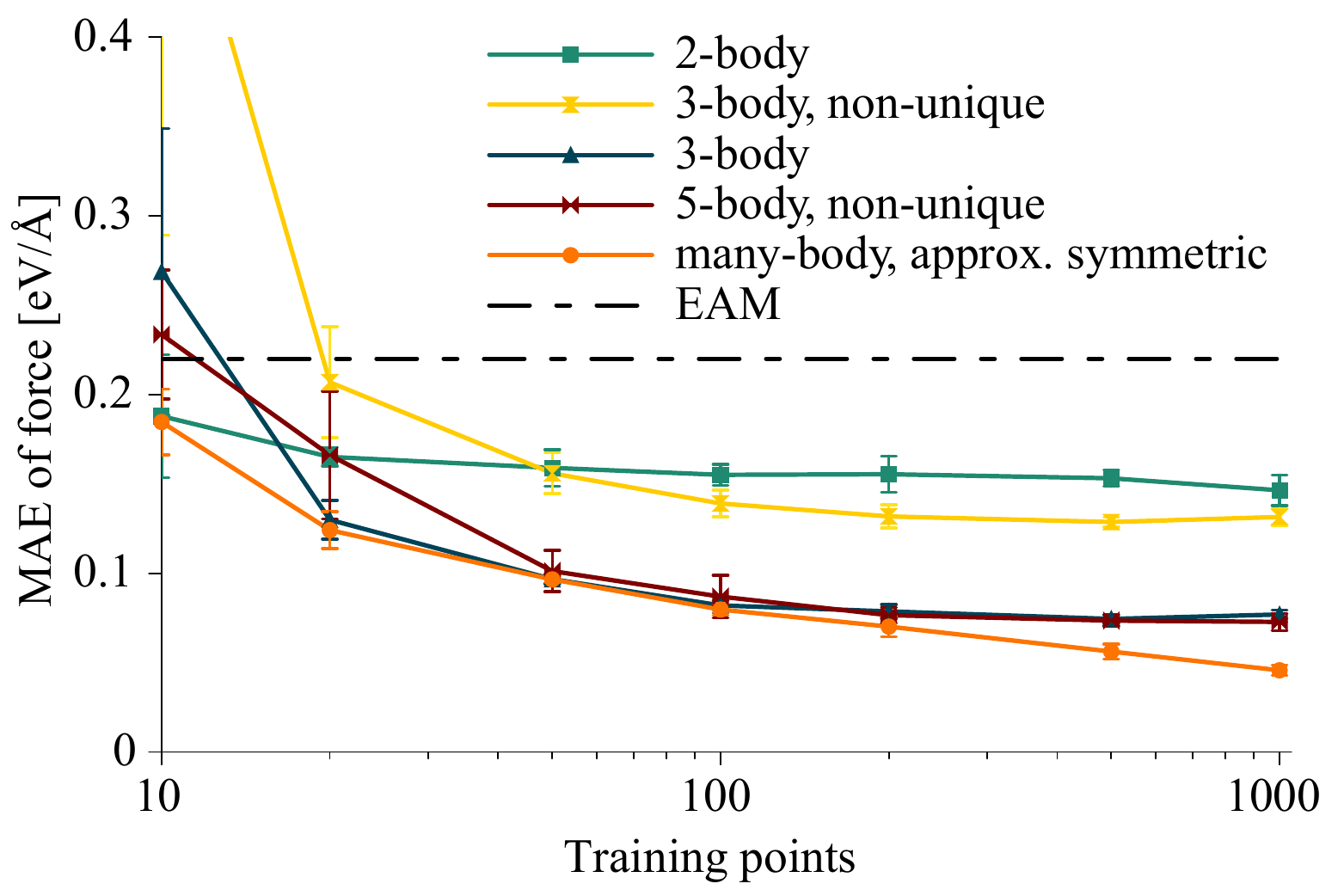}
			}
			\par\end{centering}
		\begin{centering}
			\subfloat[\label{fig:Carbon}]{\includegraphics[width=0.4\paperwidth]{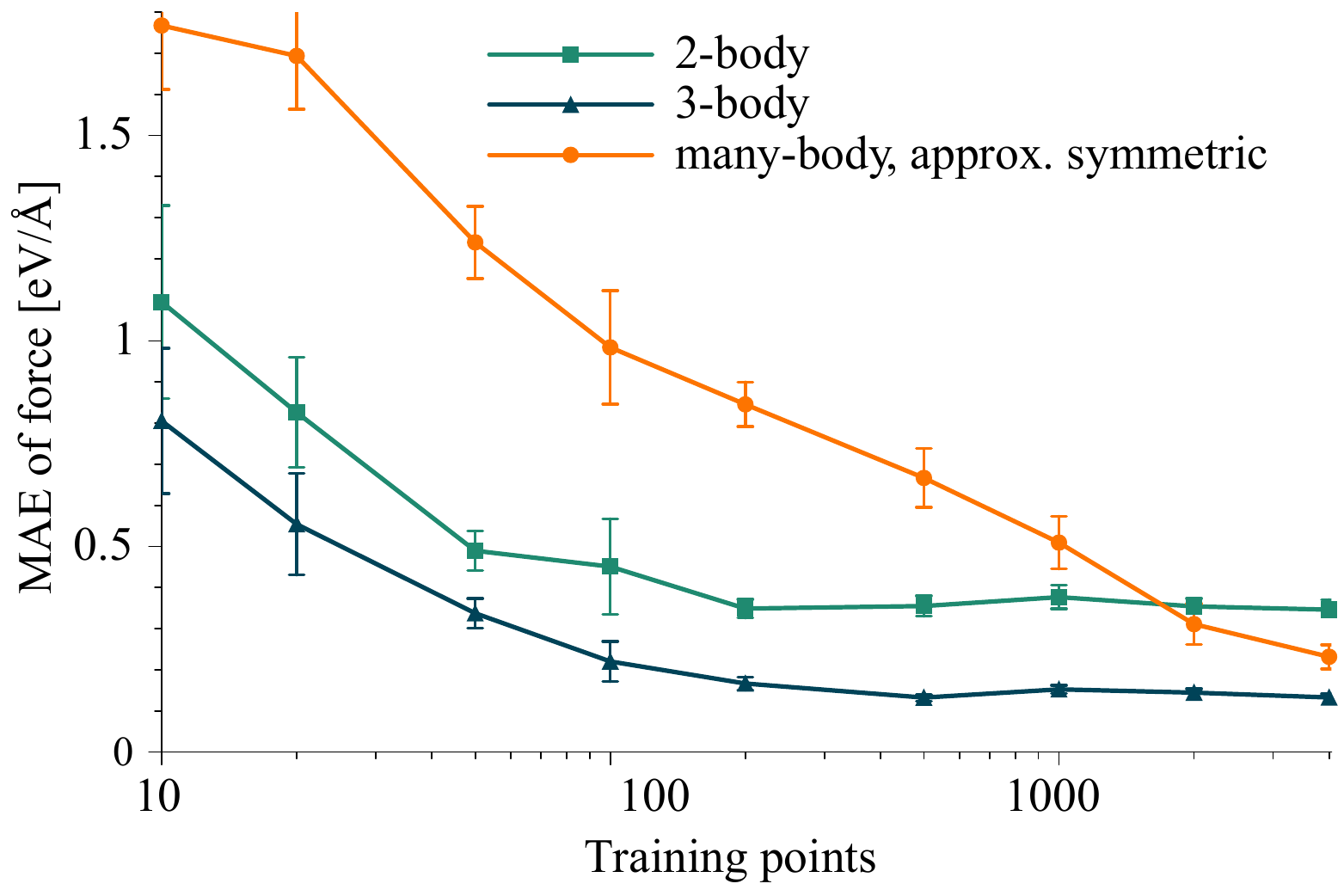}
				
			}\subfloat[\label{fig:Amorphous-Silicon}]{\includegraphics[width=0.4\paperwidth]{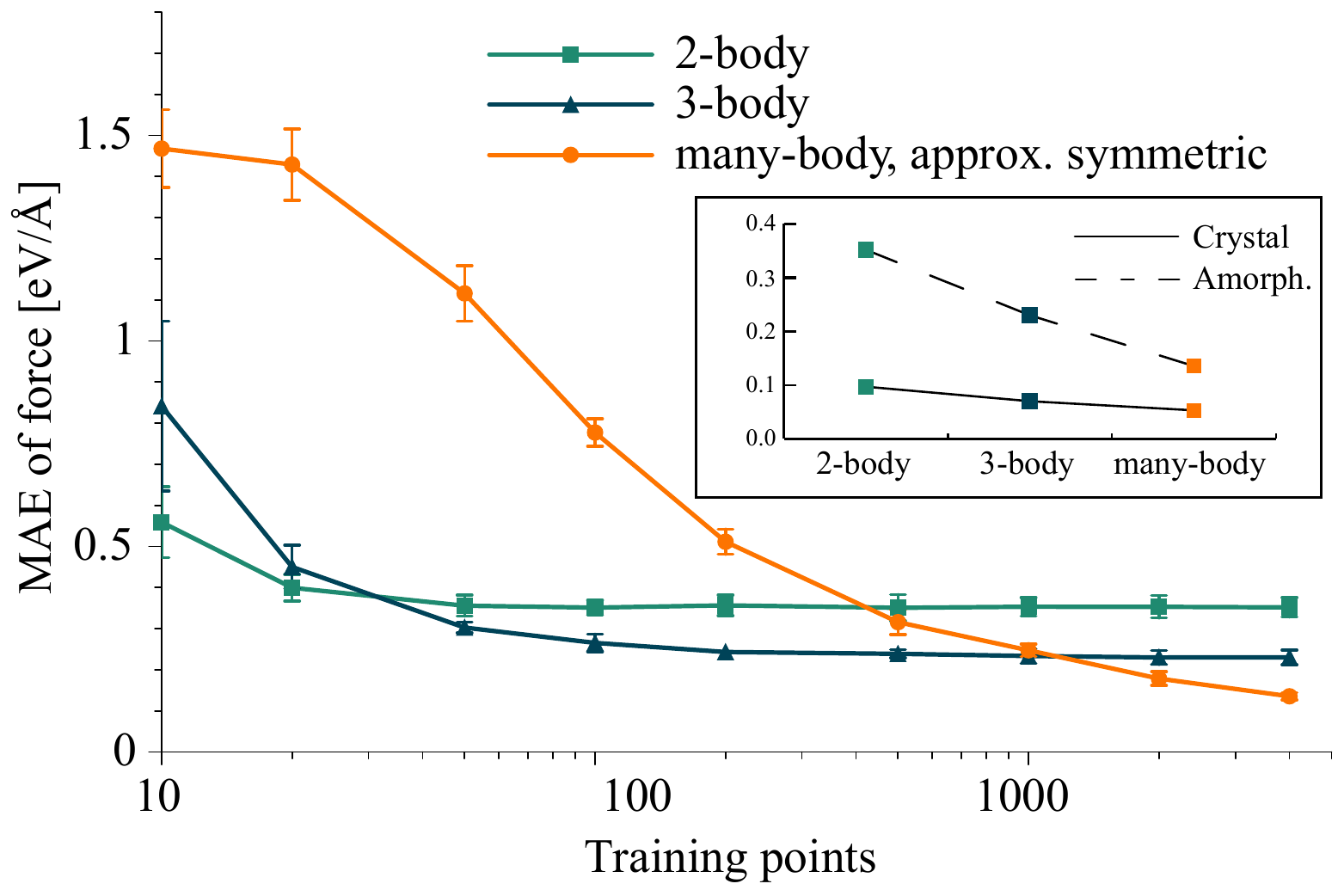}
				
			}
			\par\end{centering}
		\caption{\label{fig:MAE-on-GP} Learning curves reporting the mean generalization error (measured as the modulus of the difference between target and predicted force vectors) as a function of the training set size, for different materials and kernels of increasing order. The insets in (a) and (d) report the converged error achieved by a given kernel as a function of the kernel's order. The systems considered are: (a) Crystalline nickel, 500 K (compared to a nickel nanocluster in the inset); (b) iron with a vacancy, 500 K; (c) diamond and graphite, mixed temperatures and pressures; and (d) amorphous silicon, 650 K (compared to crystalline silicon in the inset). For extra details on the datasets and kernels used, and on the experimental methodology, see Appendixes C and D.}
	\end{figure*}
	\par\end{center}

\subsection{Optimal $\boldsymbol{n}$-kernel choice}

\begin{table}
	\begin{centering}
		%\resizebox{8cm}{!}{%
		\begin{tabular}{|c|c|c|c|c|}
			\hline 
			kernel &  order  & symm. & unique & name \tabularnewline
			\hline 
			\hline 
			$k_{2}^{s}$ & 2 & $\checked$  & $\checked$ & 2-body \tabularnewline
			\hline 
			$k_{3}^{\neg u}$ & 3 & $\checked$  & $\times$ & 3-body, non-unique \tabularnewline
			\hline 
			$k_{3}^{s}$ & 3 & $\checked$ & $\checked$ & 3-body\tabularnewline
			\hline 
			$k_{5}^{\neg u}$ & 5 & $\checked$ & $\times$ & 5-body, non-unique\tabularnewline
			\hline 
			$k_{MB}^{ds}$ & $\infty$ & $\sim$ & $\checked$ & many-body, approx. symmetric \tabularnewline
			\hline 
		\end{tabular}%}
		\par\end{centering}
	\caption{Some of the kernels presented and their properties. \label{tab: Kernels}}
\end{table}

In general, choosing a higher order $n$-body kernel will improve accuracy at the expense of speed.   
The optimal kernel choice for a given application will correspond to the best tradeoff between computational cost and representation power, which will depend on the physical system investigated.  
The properties of some of the kernels discussed above are summarized in Table \ref{tab: Kernels}, while their performance is tested on a range of materials in Figure~\ref{fig:MAE-on-GP}. 
The figure reveals some general trends. 
\mbox{2-body} kernels can be trained very quickly, as good convergence can be attained already with $\sim$100 training configurations. 
The 2-body representation is a very good descriptor for a few materials under specific conditions, while their overall accuracy is ultimately limited. 
This will yield e.g., excellent force accuracy for a close-packed bulk system like crystalline Nickel (inset (a)), and reasonable accuracy for a defected $\alpha$-Fe system whose bcc structure is however metastable if just pair potentials are used (inset (b)).   
Accuracy improves dramatically once angular information is acquired by training 3-body kernels. 
These can accurately describe forces acting on iron atoms in the bulk $\alpha$-Fe system containing a vacancy (inset (b)) and those acting on carbon atoms in both diamond and graphite (inset (c)). 
However, 3-body GPs need larger training databases. 
Also, atoms participate in many more triplets than simple bonds in their standard environments contained in the database, which will make 3-body kernels slower than 2-body ones for making predictions by GP regression.    
Both problems would extend, getting worse, to higher values of $n$, as summing over all database configurations and all feature $n$-plets in each database configuration will make GP predictions progressively slower.
However, complex materials where high-order interactions presumably play a significant role should be expected to be well described by ML-FF based on a many-body kernel. 
This is verified here in the case of amorphous Silicon (inset (d)).

Identifying the $n$ value best suited for the description of a given material system can also be done in practice by monitoring how the converged error varies as a function of the kernel order. 
Plots illustrating this behaviour are provided in insets (a) and (d) for nickel and silicon systems, respectively. 
In each plot the more complex system (a Ni cluster and an amorphous Si system, respectively) display a high accuracy gain (larger negative slope) when the kernel order is increased, while the relatively simpler cristalline Ni and Si systems show a practically constant trend on the same scale. 

Figure \ref{fig:MAE-on-GP} (b) also shows the performance of some non-unique kernels. 
As discussed above, these are options to increase the order of an input kernel avoiding the need to sum over the correspondingly higher order $n$-plets. 
Our tests indicate that the ML-FFs generated by non-unique kernels sometimes improve appreciably on the input kernels' performance: e.g., the error incurred by the 2-body kernel of Eq.~(\ref{eq: k2_k3_q}) in the Fe-vacancy system is higher than that associated with its square, the non-unique 3-body kernel of Eq.~(\ref{eq: non_unique_k3}).  
Unfortunately, but not surprisingly, the improvement can be in other cases modest or nearly absent, as exemplified by comparing the errors associated with the 3-body kernel and its square -the non-unique 5-body kernel-, in the same system. 

Overall, the analysis of Figure \ref{fig:MAE-on-GP} suggests that an  optimal kernel can be chosen by comparing the learning curves of the various $n$-body kernels and the many-body kernel over the available QM database: the comparison will reveal the simplest (most informative, lowest $n$) description that is still compatible with the error level deemed acceptable in the simulation.

Trading transferability for accuracy by training the kernels on a QM database appropriately tailored for the target system (e.g., restricted to just bulk or simply-defected system configurations sampled at the relevant temperatures 
as done in the Ni and Fe-systems of Figure \ref{fig:MAE-on-GP}) will enable surprisingly good accuracy even for low $n$ values. 
This should be expected to systematically improve on the accuracy performance of classical potentials involving non-linear parameter fitting, as exemplified by comparing the errors associated with $n$-body kernel models and the average errors of state-of-the-art embedded atom model (EAM) P-FFs \cite{Mishin:2004bh,Mendelev:2003co} (insets (a) and (b)).
The next section further explores the performance of GP-based force prediction, to address the final issue of what execution speed can be expected for ML-based force fields, once the optimally accurate choice of kernel has been made.

\section{Mapped Force Fields (M-FFs)}

\begin{figure}
	\includegraphics[width=1.0\columnwidth]{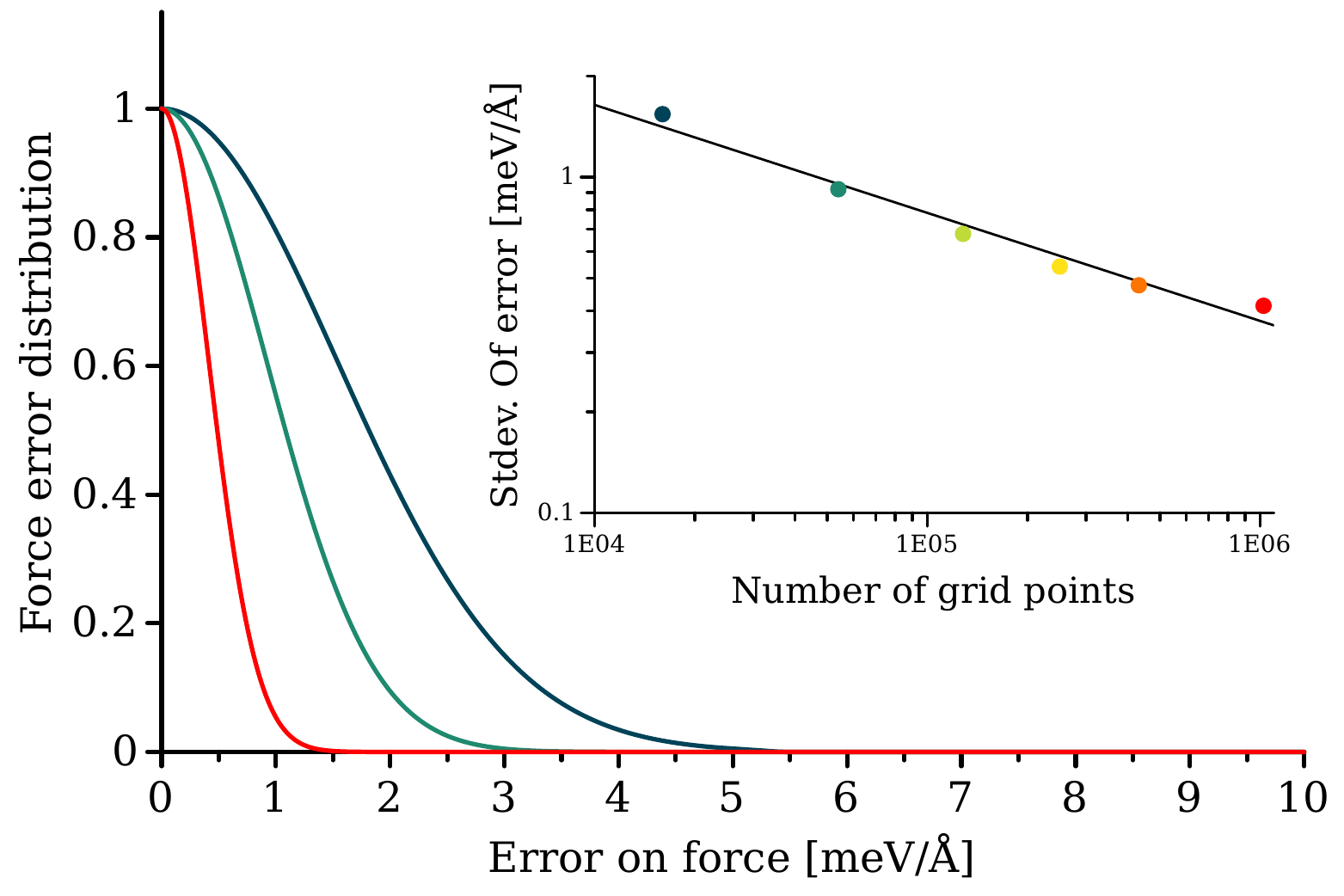}
	
	\caption{ 
nser grids. Inset: standard deviation of the distributions as a function of the number of interpolation grid points, on a log-log scale (each distribution in the main panel corresponds to a dot of same color in the insert).  
		\label{fig: Mapping error}}
\end{figure}

Once a GP kernel is recognized as being $n$-body, it automatically defines an $n$-body force field corresponding to it, for any given choice of training set.
This will be an $n$-body function of atomic positions satisfying Eq.~\eqref{eq: nBody_ker_def}, whose values {\em can}  be computed by GP regression sums over the training set as done by standard ML-FF implementations, but  {\em do not have to} be computed this way. 
In particular, the execution speed of a machine learning-derived  $n$-body force field might be expected to depend on its order $n$  (e.g., it will involve sums over all atomic triplets, like any  3-body P-FF, if $n$=3), but should otherwise be independent of the training set size. 
It should therefore be possible to construct a mapping procedure yielding a machine learning-derived, nonparametric force field (an efficient \q{M-FF}) that allows a very significant speed-up over calculating forces by direct GP regression.    
We note that non-unique kernels obtained as powers of $n'$-body input kernels exceed their reference $n'$-body feature space and thus could not be similarly sped up by mapping their predictions onto an M-FF of equal order $n'$, while mapping onto an M-FF of the higher output order $n$ would still be feasible.
For convenience, we will analyze a 3-body kernel case, show that a 3-body GP exactly corresponds to a classical 3-body M-FF, and show how the mapping yielding the M-FF can be carried out in this case, using a 3D-spline approximator.  
The generalization to any order $n$ is straightforward provided that a good approximator can be identified and implemented.   
We begin by inserting the general form of a 3-body kernel (Eq.~(\ref{eq: kn_q})) into the GP prediction expression (Eq.~(\ref{eq: GP_pred})), to obtain

\begin{align}
\varepsilon(\rho) & =\sum_{d=1}^{N}\sum_{\substack{i_1,i_2\in\rho\\
j_1,j_2\in\rho_{d}
}
}\tilde{k}_{3}(\mathbf{q}_{i_1,i_2},\mathbf{q}_{j_1,j_2}^{d})\alpha_{d}.
\label{eq: epsilon(rho)}
\end{align}
Inverting the order of the sums over the database and atoms in the target configurations yields a general expression
for the 3-body potential:
\begin{align}
\varepsilon(\rho) & =\sum_{i_1,i_2\in\rho}\tilde{\varepsilon}(\mathbf{q}_{i_1,i_2})\label{eq: epsilon(rho)_2} \\
\tilde{\varepsilon}(\mathbf{q}_{i_1,i_2}) & =\sum_{d=1}^{N}\sum_{j_1,j_2\in\rho_{d}}\tilde{k}_{3}(\mathbf{q}_{i_1,i_2},\mathbf{q}_{j_1,j_2}^{d})\alpha_{d}.\label{eq: 3-body_pot}
\end{align}

Eq.~(\ref{eq: epsilon(rho)_2}) reveals that the GP implicitly defines the local energy of a configuration as a sum over all triplets containing the central atom, where the function $\tilde{\varepsilon}$ represents the energy associated with each triplet in the physical system. 
The triplet energy is calculated by three nested sums, one over the $N$ database entries and two running over the $M$ atoms of each database configuration ($M$ may slightly vary over configurations, but can be assumed to be constant for the present purpose).  
The computational cost of a single evaluation of the triplet energy \eqref{eq: 3-body_pot} scales consequently as $\mathcal{O}(NM^{2})$. 
Clearly, improving the GP prediction accuracy by increasing $N$ and $M$ will make the prediction slower.  
However, such a computational burden can be avoided, bringing the complexity of the triplet energy calculation \eqref{eq: 3-body_pot} to $\mathcal{O}(1)$.

\begin{figure}
	\includegraphics[width=1.0\columnwidth]{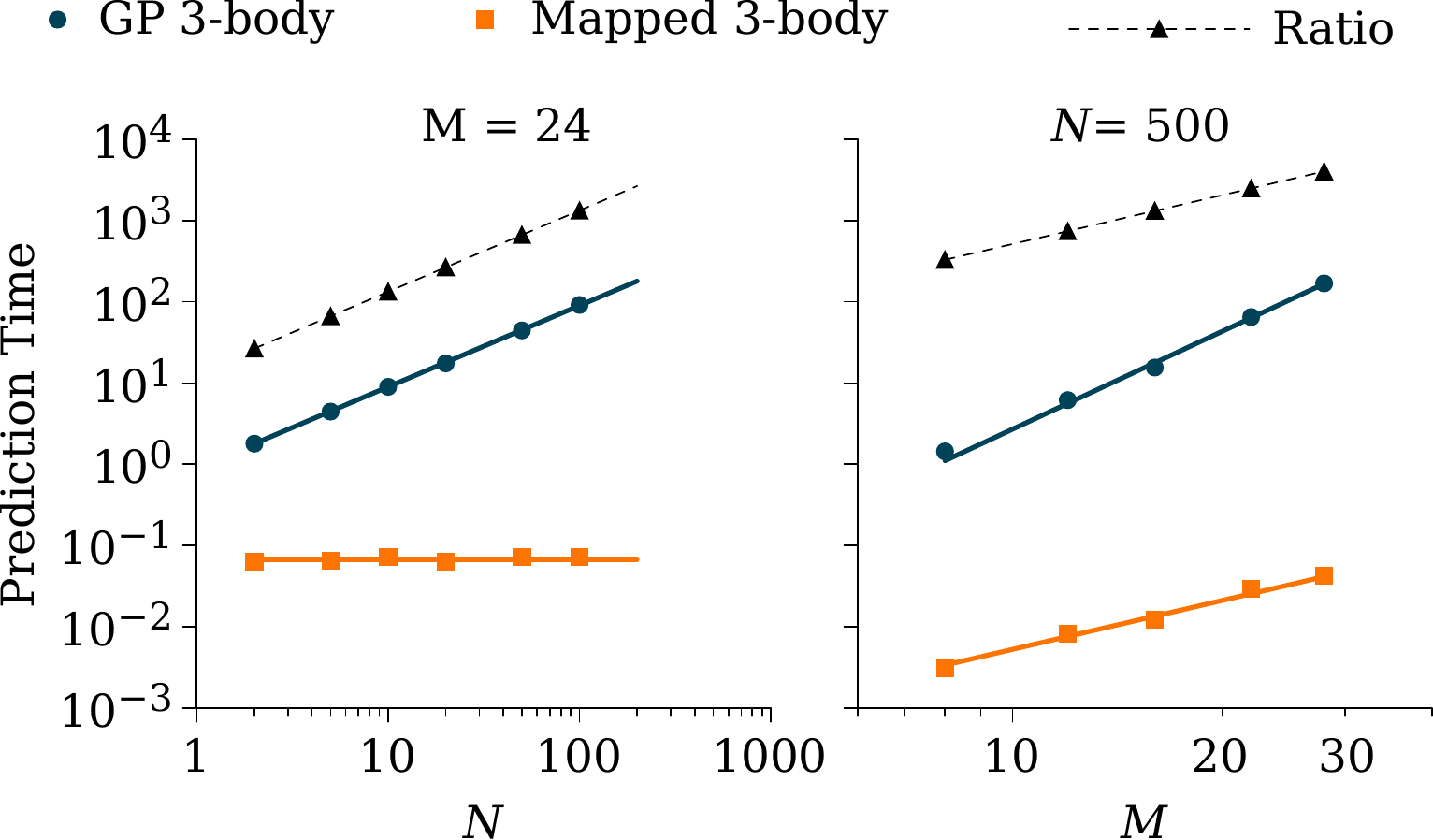}
	\caption{ Computational cost of evaluating the 3-body energy (Eq.~(\ref{eq: epsilon(rho)_2})) as a function of the database size $N$ and the number of atoms $M$ located within the cutoff radius. 
		Left: time taken ($\rm{s}$) for a single energy prediction using the GP (blue dots and solid line) and the mapped potential (orange dots and solid line), as a function of $N$, for $M$=24. The speed-up ratio is also provided as a dotted line.
		Right: scaling of the same quantities as a function of  $M$ for a  training set of $N = 500$ configurations. 
		\label{fig: Remapping speedup}}
\end{figure}

Since the triplet energy $\tilde{\varepsilon}$ is a function of just three variables (the effective symmetry-invariant degrees of freedom associated with three particles in three dimensions), we can calculate and store its values on an appropriately distributed grid of points within its domain. 
This procedure effectively maps the GP predictions on the relevant 3-body feature space: once completed, the value of the triplet energy at any new target point can be calculated via a local interpolation, using just a subset of nearest tabulated grid points. 
If the number of grid points $N_{g}$ is made sufficiently high, the mapped function will be essentially identical to the original one but, by virtue of the locality of the interpolation, the cost of evaluating it will not depend on $N_{g}$.

The 3-body configuration energy of Eq.~\eqref{eq: epsilon(rho)_2} also includes 2-body contributions coming from the terms in the sum for which the indices $i_1$ and $i_2$ are equal. 
When $i_1 = i_2 = i$ the term $\varepsilon(\mathbf{q}_{i,i})$ can be interpreted as the pairwise energy associated with the central atom and atom $i$. 
The term can consequently be mapped onto a 1D 2-body feature space whose coordinate is the single independent component of the $\mathbf{q}_{i,i}$ feature vector, typically the distance between atom $i$ and the central atom. 
In the same way, an $n$-body kernel naturally defines a set of $n$-body energy terms of order comprised between $2$ and $n$, depending on the number of repeated indices.

\begin{figure}
	\includegraphics[width=1.\columnwidth]{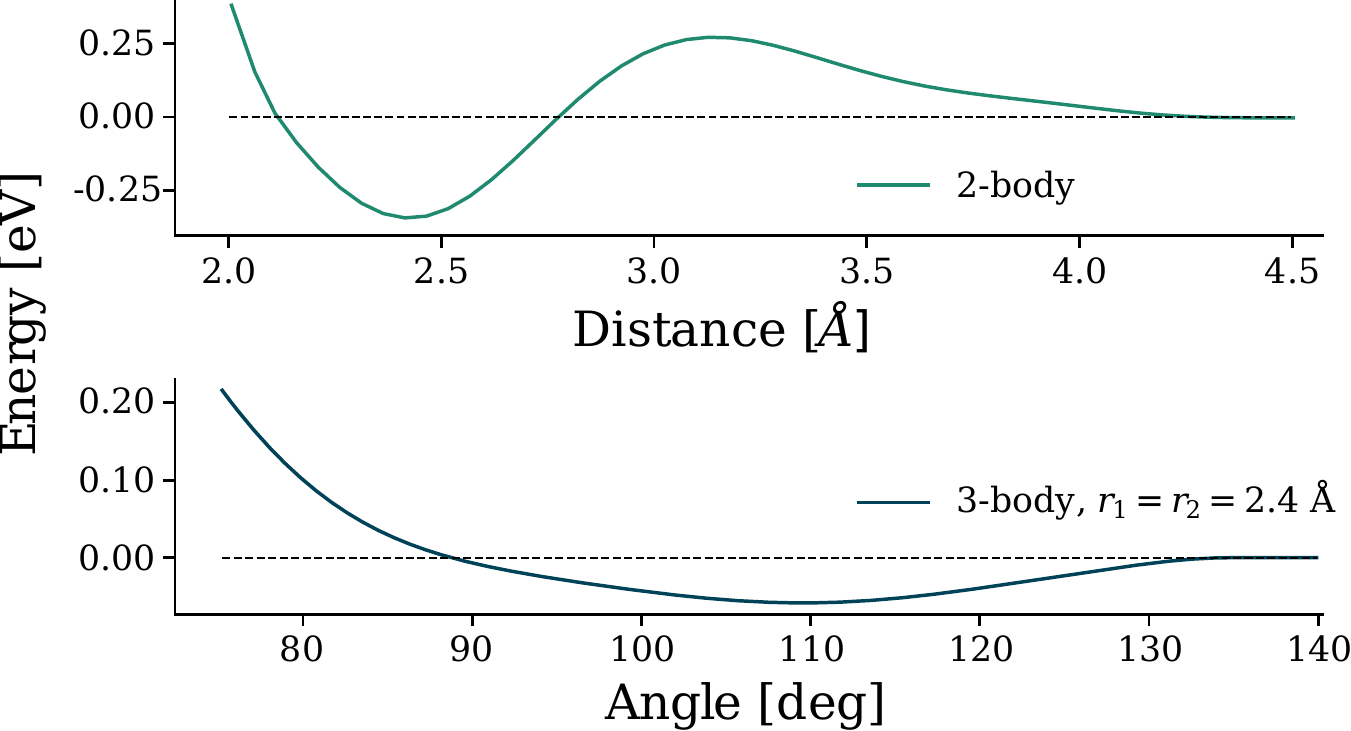}
	\caption{Energy profiles of the 3-body M-FF, trained for the a-Si system at 650K. Upper panel: 2-body interaction term. Lower panel: 3-body interaction energy for an atomic triplet, angular dependence when the two distances from the central atoms are both equal to 2.4\AA.}
	\label{fig: Si_pot}
\end{figure}

Figure \ref{fig: Mapping error} shows the convergence of the mapped forces derived from the 3-body kernel in Eq.~\eqref{3bodyk} for a database of DFTB atomic forces for the a-Si system. 
The interpolation is carried out using a 3D cubic spline for different 3D mesh sizes.
Comparison with the reference forces produced by the GP allows to generate, for each mesh size, the distribution of the absolute deviation of the force components from their GP-predicted values.   
The standard deviation of the interpolation error distribution is shown in the insert on a log-log scale, as a function of $N_{g}$. 
Depending on the specific reference implementation, the speed-up in calculating the local energy (Eq.~\eqref{eq: epsilon(rho)_2}) provided by the mapping procedure can vary widely, while it will always grow linearly with $N$ and quadratically with $M$ (see Figure \ref{fig: Remapping speedup}), and it will be always substantial: in typical testing scenarios we found this to be of the order of $10^{3}-10^{4}$. 

An M-FF example, obtained for a-Si with $n=3$ is shown in Figure \ref{fig: Si_pot}. 
As the energy profile is not prescribed by any particular functional form, it is free to optimally adapt to the information contained in the QM training set, to best reproduce the quantum interactions that produced it. 
Figure \ref{fig: Si_pot} contains some expected features e.g., a radial minimum at about $r \simeq2.4\text{\AA}$  in the 2-body section (upper panel), the corresponding angular minimum at $\theta_0\simeq110^{\circ}$ (lower panel), which is approximately equal to the $sp^3$ hybridization angle of 109.47$^{\circ}$, and rapid growth for small radii (upper panel) and angles (lower panel). 
Less intuitive features are also visible, which however contribute to the best representation of the bulk system's interactions that a 3-body expansion can achieve for the given database. 
An example is the shallow maximum in the 2-body section at $r\simeq3.1\text{\AA}$, which would of course disappear if we fitted our model on QM forces calculated for a Si dimer,  that do not contain a hump. 
The resulting Si force field, appropriate for a Si dimer,  would however inevitably reproduce the QM bulk interactions less accurately. 
More generally, training on the aggregate dataset could be a sensible compromise, producing a more transferable, but locally less accurate force field.  
%
%Alternatively, an efficient strategy for simulating complex systems with space- and time-varying bonding nature might involve using (concurrently across the system, and at any given time) the locally optimal choice of low-order M-FF.   

%
%
\section{Concluding remarks}

The results presented in this work exemplify how physical priors built in the GP kernels restrict their descriptiveness, while improving their convergence speed as a function of training dataset size. 
This provides a framework to optimise efficiency. 
Comparing the performance of $n$-body kernels allows us to identify the lowest order $n$ that is compatible with the required accuracy as a consequence of the physical properties of the target system. 
As a result, accuracy can in each case be achieved using the most efficient kernel e.g., a 2-body kernel for bulk Ni, or a 3-body kernel for carbon and graphite, for a $\sim$ 0.1~eV/$\text{\AA}$ target force accuracy, see Figure \ref{fig:MAE-on-GP}. 
As should be reasonably expected, relying on low-dimensional feature spaces will limit the maximum achievable accuracy if higher-order interactions, missing in the representation, occur in the target system.  

On the other hand, we find that once the optimal order $n$ has been identified and the $n$-kernel has been trained (whichever its form e.g., whether defined as a function of invariant  descriptors as in (\ref{eq: kn_q}), or constructed as a power of such a function, or derived as an analytic integral over rotations using Eqs. (\ref{eq: kn_decomposition}-\ref{eq: JAMES_2})), it becomes possible to map its prediction on the appropriate $n$-dimensional domain, and thus generate a $n$-body M-FF machine-learned atomic force field that predicts atomic forces at essentially the same computational cost of a classical $n$-body P-FF parametric force field. 
The GP predictions allow for a natural intrinsic measure of uncertainty -the GP predicted variance-, and the same mapping procedure used for the former can also be applied to the latter.
Thus, like their ML-FF counterparts, and unlike P-FFs, M-FFs offer a tool which could be used to monitor whether any extrapolation is taking place that might involve large prediction errors. 

In general, our results suggest a possible three-step procedure to build fast nonparametric M-FFs whenever 
a series of kernels $k_n$ can be defined with progressively higher complexity/descriptive power and well-defined feature spaces with $n$-dependent dimensionality.
However the series is constructed (and whether or not it converges to a universal descriptor) this will involve   
(i) GP training of $n$-kernels for different values of $n$, in each case using as many database entries relevant to the target system as needed to achieve convergence of the $n$-dependent prediction error; 
(ii) identification of the optimal order $n$, yielding the simplest viable description of the system's interactions - this could be e.g., the minimal value of $n$ compatible with the target accuracy or, within a GP statistical learning framework, the result of a Bayesian model selection procedure \cite{Williams:2006vz};  
(iii) mapping of the resulting (GP-predicted) ML-FF onto an efficient M-FF, using a suitably fast approximator function defined over the relevant feature space.  

A major limitation of the M-FFs obtained this way is that, similar to P-FFs, each of them can be used only in \q{interpolation mode}, that is when the target configurations are all well represented in the fixed database used. 
This is not the case in molecular dynamics simulations potentially revealing new chemical reaction paths, or whenever online learning or the use of dynamically-adjusted database subsets are necessary to avoid unvalidated extrapolations and maximise efficiency.  
In such cases, \q{learn on the fly} (LOTF) algorithms can be deployed, which have the ability to incorporate novel QM data into the database used for  force prediction. 
In such schemes, the new data are either developed at runtime by new QM calculations, or they are adaptably retrieved as the most relevant subset of a much larger available QM database \cite{Li:2015eb}. 
The availability of an array of $n$-body kernels is very useful for this class of algorithms, which provides further motivation for their development.   
In particular, distributing the use of $n$-body kernels non-uniformly in both space and time along the system's trajectory has the potential to provide an optimally efficient approach to accurate MD simulations using the LOTF scheme. 
Finally, while the complication of continuously mapping the GP predictions to reflect a dynamically updated training database makes on the fly M-FF generation a less attractive option, a strategy to produce the MD trajectory with classical force field efficiency might involve using (concurrently across the system, and at any given time) a locally optimal choice built from a comprehensive set of pre-computed low-order M-FFs.

\section*{Acknowledgements}

The authors acknowledge funding by the Engineering and Physical Sciences
Research Council (EPSRC) through the Centre for Doctoral Training \q{Cross Disciplinary Approaches to Non-Equilibrium Systems} (CANES, Grant No. EP/L015854/1) and by the Office of Naval Research Global (ONRG Award No. N62909-15-1-N079). ADV acknowledges further support by the EPSRC HEmS Grant No. EP/L014742/1 and by the European Union\textquoteright s Horizon 2020 research and innovation program (Grant No. 676580, The NOMAD Laboratory, a European Centre of Excellence). We are grateful to the UK Materials and Molecular Modelling Hub for computational resources, which is partially funded by EPSRC (EP/P020194/1). We furthermore thank Gábor Csányi, from the Engineering Department, University of Cambridge, for the Carbon database and Samuel Huberman, from the MIT Nano-Engineering Group for the initial geometry used in the a-Si simulations. Finally, we want to thank Ádám Fekete for stimulating discussions and precious technical help.

\bibliography{nK,CK,CZ}

\section*{APPENDIX}

\subsection{Kernel order by explicit differentiation}

We first prove that the kernel given in Eq.~(\ref{eq: k2}) is 2-body in the sense of Eq.~(\ref{eq: nBody_ker_def}). 
For this it is sufficient to show that its second derivative with respect to the relative position of two different atoms of the target configuration $\rho$ always vanishes. 
The first derivative is
\begin{align*}
\frac{\partial k_{2}(\rho, \rho ')}{\partial\mathbf{r}_{i_1}} & =\sum_{ij}\frac{\partial}{\partial\mathbf{r}_{i_1}}\mathrm{e}^{-\|\mathbf{r}_{i}-\mathbf{r}_{j}'\|^{2}/4\sigma^{2}}\\
& =\sum_{ij}\mathrm{e}^{-\|\mathbf{r}_{i}-\mathbf{r}_{j}'\|^{2}/4\sigma^{2}}\frac{(\mathbf{r}_{i}-\mathbf{r}_{j}')}{2\sigma^{2}}\delta_{ii_1}\\
 & =\sum_{j}\mathrm{e}^{-\|\mathbf{r}_{i_1}-\mathbf{r}_{j}'\|^{2}/4\sigma^{2}}\frac{(\mathbf{r}_{i_1}-\mathbf{r}_{j}')}{2\sigma^{2}}.
\end{align*}
This depends only on the atom located at $\mathbf{r}_{i_1}$ of the configuration $\rho$. 
Thus, differentiating with respect to the relative position $\mathbf{r}_{i_2}$ of any other atom of the configuration gives the relation in Eq.~(\ref{eq: nBody_ker_def}) for 2-body kernels: 
\[
\frac{\partial^{2}k_{2}(\rho,\rho')}{\partial\mathbf{r}_{i_1}\partial\mathbf{r}_{i_2}}=0 . 
\]

We next show that the kernel defined in Eq.~(\ref{eq: kn}) is an $n$-body in the sense of Eq.~(\ref{eq: nBody_ker_def}).
This follows naturally from the result above, given that $k_n$ is defined as $k_n = k_2^{n-1}$. 
We can thus write down its first derivative as
\[
\frac{\partial k_{n}}{\partial\mathbf{r}_{i_1}}=(n-1)k_{2}^{n-2}\frac{\partial k_{2}}{\partial\mathbf{r}_{i_1}}.
\]
Since the second derivative of $k_2$ is null, the second derivative of $k_n$ is simply
\begin{align*}
\frac{\partial^{2}k_{n}}{\partial\mathbf{r}_{i_1}\partial\mathbf{r}_{i_2}} 
& = (n-2)(n-1)k_{2}^{n-3}\frac{\partial k_{2}}{\partial\mathbf{r}_{i_1}}\frac{\partial k_{2}}{\partial\mathbf{r}_{i_2}}
\end{align*}
and after $n-1$ derivations we similarly obtain
\begin{align*}
\frac{\partial^{2}k_{2}^{n-1}}{\partial\mathbf{r}_{i_1}\cdots\partial\mathbf{r}_{i_n}} & =(n-1)! \,k_{2}^{0} \,\frac{\partial k_{2}}{\partial\mathbf{r}_{i_1}}\dots\frac{\partial k_{2}}{\partial\mathbf{r}_{i_{n-1}}}.\\
\end{align*}
Since $k_{2}^{0}=1$, the final derivative with respect to 
the $n_{th}$ particle position $\mathbf{r}_{i_n}$ is zero as required by Eq.~(\ref{eq: nBody_ker_def}). 

\subsection{1D $\boldsymbol{n'}$-body model}
To test the ideas behind the $n$-body kernels, we used a 1D $n'$-particle model reference system where a (\q{central}) particle is kept fixed at the coordinate axis origin (consistent with the local configuration convention of Eq.~(\ref{eq: Gauss_exp})). 
The energy of the central particle is defined as
\begin{equation*}
f = \sum_{i_1\dots i_{{n'}-1}} J \, x_{i_1} \dots x_{i_{{n'}-1}}
\end{equation*}
where $\{x_{i_p}\}_{p=1}^{{n'}-1}$ are the relative positions of ${n'}-1$ particles, and  $J$ is an interaction constant.

To generate Figure \ref{fig: simple_test} a large set of configurations was generated by uniformly and independently sampling each relative position $x_{i_p}$ within the range $(-0.5, 0.5)$. The energy of the central particle of each configuration was then given by the above equation, with the interaction constant $J$ set to $0.5$. In order to analyse the converged properties of the $n$-body kernels presented, large training sets ($N = 1000$) were used. 
\subsection{Databases details}

The bulk Ni and Fe databases were obtained from simulations using a $4\times4\times4$ periodically repeated unit cell, modelling the electronic exchange and correlation interactions via the PBE/GGA approximation \cite{Perdew:1996iq}, and controlling the temperature (set at $\rm{500K}$) by means of a weakly-coupled Langevin thermostat (the DFT trajectories are available from the KCL research data management system at the link http://doi.org/10.18742/RDM01-92). 
The C database comprises bulk diamond and AB- and ABC-stacked graphene layer structures.
These structures were obtained from DFT simulations at varying temperatures and pressures, using a fixed 3$\times$3$\times$2 periodic cell geometry for graphite, and simulation cells ranging from 1$\times$1$\times$1 to 2$\times$2$\times$2 unitary cells for diamond, the relative DFT trajectories can be found in the \q{libAtoms} data repository via the following link http://www.libatoms.org/Home/DataRepository.
Crystalline and amorphous Si database was obtained from a microcanonical DFTB 64-atom simulation carried out in periodic boundaries, with average kinetic energy corresponding to a temperature of $T = 650K$.
The results presented for the Ni cluster are reported from Ref.~\cite{Zeni:2018to} and correspond to constant temperature DFT MD runs ($T = 300K$)  of a particular $\text{Ni}_{19}$ geometry (named \q{4HCP} in the article).
The radial cutoffs used to create the local environments for the four elements considered are: 4.0 Å (Ni), 4.45 Å (Fe), 3.7 Å (C) and 4.5 Å (Si).

\subsection{Details on the kernels used and on the experimental methodology}
 
 All energy kernels presented in the work can be used to learn/predict forces after generating a standard derivative kernel (Ref. \cite{Williams:2006vz}, section 9.4, also cf. Section II A of the main text.) 
 In particular, for each scalar energy kernel $k$ a matrix-valued force kernel $\mathbf{K}$ can be readily obtained by double differentiation with respect to the positions ($\mathbf{r}_0$ and $\mathbf{r}'_0$) of the central atoms in the target and database configurations $\rho$ and $\rho'$: 
 \begin{equation*}
 \mathbf{K}(\rho, \rho') = \dfrac{\partial^2 k(\rho, \rho')}{\partial \mathbf{r}_{0} \partial \mathbf{r}_{0}^{\prime \rm{T}}}.
 \label{eq:3b_force_kernel}
 \end{equation*}
 
The kernels $\tilde{k}_2$ and $\tilde{k}_3$ (Eqs. (\ref{eq: k2_k3_q},\ref{3bodyk})) were chosen as simple squared exponentials in the tests shown. 
Noting as $\mathbf{q}$ (or $q$) the vector (or scalar) containing the effective degrees of freedom of the atomic $n$-plet considered (see Eq. (\ref{eq: kn_q})), the two kernels read:
 \begin{align*}
 \tilde{k}_2(q_i,q_j') & = \mathrm{e}^{-(q_i-q'_j)^2/2\sigma^2} \\
 \tilde{k}_3(\mathbf{q}_{i_1,i_2},\mathbf{q}'_{j_1, j_2})	& = \sum_{\mathbf{P} \in \mathcal{P}_c} \mathrm{e}^{-\|\mathbf{q}_{i_1, i_2}- \mathbf{P} \mathbf{q}'_{j_1,j_2}\|^2/2\sigma^2} ,
 \end{align*}
 where $\mathcal{P}_c$  (|$\mathcal{P}_c| = 3$) is the set of cyclic permutations of three elements. 
 Summing over the permutation group is needed to guarantee permutation symmetry of the energy.
As discussed in the main text, the exact form of these low-order kernels is not essential as the large database limit is quickly reached. 

The many-body force kernel referred to in Fig. \ref{fig:MAE-on-GP} was built as a covariant discrete summation of the many-body energy base-kernel \eqref{eq: k_SE}  over the $O_{48}$ crystallographic point group, using the procedure of Ref. \cite{Glielmo:2017dj}.   
This procedure yields an approximation to the full covariant integral of the many-body kernel \eqref{eq: k_SE} given in Eq.~\eqref{eq: k_cov}.

In order to obtain Figure 5, repeated (randomised) realisations of the same learning curves were performed. The points (and error bars) plotted are the means (and standard deviations) of the generated data . The kernel hyperparameters were independently optimised  by cross validation for each dataset.
\end{document}